\documentclass[fleqn,usenatbib]{mnras}
\usepackage{newtxtext,newtxmath}
\usepackage[T1]{fontenc}
\usepackage{ae,aecompl}
\usepackage{times}
\usepackage{comment}

\usepackage{graphicx}	
\graphicspath{{/Users/chessipearce/Documents/PhD/Paper2/plots/}}
\usepackage{amsmath}	
\usepackage{amssymb}	
\usepackage{comment}
\usepackage{etoolbox}
\makeatletter
\patchcmd\@combinedblfloats{\box\@outputbox}{\unvbox\@outputbox}{}{%
	\errmessage{\noexpand\@combinedblfloats could not be patched}%
}%
\makeatother

\newcommand{\msol}{M$_{\odot}$}

\title[C-EAGLE metallicity evolution]{Redshift evolution of the hot intracluster gas metallicity in the C-EAGLE cluster simulations}

\author[F. A. Pearce et al.]{Francesca A. Pearce$^{1}$,\thanks{E-mail: francesca.pearce@manchester.ac.uk}
Scott T. Kay$^{1}$, David J. Barnes$^{2}$, Yannick M. Bah\'{e}$^{3}$, and \newauthor Richard G. Bower$^{4}$
\\
$^{1}$Jodrell Bank Centre for Astrophysics, Department of Physics and Astronomy, The University of Manchester, Manchester M13 9PL, UK\\
$^{2}$Departmentof Physics, Kavli Institute for Astrophysics and Space Research, Massachusetts Institute of Technology, Cambridge, MA 02139, USA\\
$^{3}$Leiden Observatory, Leiden University, PO Box 9513, 2300 RA Leiden, the Netherlands\\
$^{4}$Institute for Computational Cosmology, Department of Physics, University of Durham, South Road, Durham DH1 3LE, UK
}

\date{Accepted XXX. Received YYY; in original form ZZZ}

\pubyear{2019}

\begin{document}
\label{firstpage}
\pagerange{\pageref{firstpage}--\pageref{lastpage}}
\maketitle

\begin{abstract}
	The abundance and distribution of metals in galaxy clusters contains valuable information about their chemical history and evolution. By looking at how metallicity evolves with redshift, it is possible to constrain the different metal production channels. We use the C-EAGLE clusters, a sample of 30 high resolution ($m_{\mathrm{gas}} \simeq 1.8\times 10^{6}$~M$_{\odot}$) cluster zoom simulations, to investigate the redshift evolution of metallicity, with particular focus on the cluster outskirts. The early enrichment model, in which the majority of metals are produced in the core of cluster progenitors at high redshift, suggests that metals in cluster outskirts have not significantly evolved since $z=2$. With the C-EAGLE sample, we find reasonable agreement with the early enrichment model as there is very little scatter in the metallicity abundance at large radius across the whole sample, out to at least $z=2$. The exception is Fe for which the radial dependence of metallicity was found to evolve at low redshift as a result of being mainly produced by Type Ia supernovae, which are more likely to be formed at later times than core-collapse supernovae. We also found considerable redshift evolution of metal abundances in the cores of the C-EAGLE clusters which has not been seen in other simulations or observation based metallicity studies. Since we find this evolution to be driven by accretion of low metallicity gas, it suggests that the interaction between outflowing, AGN heated material and the surrounding gas is important for determining the core abundances in clusters.
\end{abstract}




\section{Introduction}

Galaxy clusters are the largest virialised objects in the present day Universe, formed from the collapse of the rarest and largest fluctuations in the primordial density field. Baryons account for around 15 per cent of the mass of a cluster, with the rest coming from dark matter which cannot be observed directly, and is inferred through its gravitational effect on other observable matter. The majority of baryons in clusters are found in the intracluster medium (ICM, e.g \citealt{Renzini2014}), a very hot $T \sim 10^{8}$~K plasma which contains a wealth of information on cluster formation and evolution. The ICM can be probed directly using X-ray observations, from which the density and thermal properties of the ICM can be derived, and subsequently the density. The abundance and distribution of metals in a cluster can also be found from the analysis of X-ray line emissions, which can be used to place constraints on the different enrichment channels and also give insight into the chemical history and evolution of clusters (see \citealt{Mernier2018} for a recent review of current observations and constraints).

The study of metal abundance in the ICM started in the 1970's with the first detection of the iron (Fe) K-shell emission line complex by Ariel-V \citep{Mitchell1976} and OSO-8 \citep{Serlemitsos1977}. The \textit{ASCA} observatory helped to advance the field with more robust measurements of the iron abundance in nearby galaxy clusters (e.g. \citealt{Mushotzky1996, Fukazawa1998}). It also provided the first measurements of other previously undetected metals; oxygen (O), neon (Ne), magnesium (Mg), silicon (Si), sulphur (S), argon (Ar), calcium (Ca) and nickel (Ni). The latest advancements in the study of metals have come from the latest generation of X-ray observatories, namely \textit{Chandra}, \textit{XMM-Newton} and \textit{Suzaku}. These missions allowed the spatial distribution of metals to be measured in greater detail, and were able to help characterise the redshift evolution of metallicity. However, to date the best measurements of metallicity have been of the core of the Perseus cluster by the Hitomi spacecraft \citep{Hitomi2017}. In contrast to previous analyses, Hitomi found the ICM abundance in Perseus to be fully consistent with the proto-solar abundance of \cite{Lodders2009}, something which cannot easily be reproduced using linear combinations of existing supernova nucleosynthesis calculations \citep{Simionescu2019}. 

The chemical enrichment of clusters is driven by stars, with many different astrophysical processes such as active galactic nuclei (AGN) outbursts, turbulent diffusion and galactic winds contributing to the distribution of metals throughout the ICM (e.g. \citealt{Schindler2008}). The majority of heavy elements, specifically Fe and Ni, are produced through Type Ia supernovae (SNe), explosions of white dwarf stars in binary systems. The collapse of massive, short-lived stars in Type II SNe produces O, Ne and Mg, and some intermediate-mass elements such as Si, S, Ar and Ca. Other elements such as C, N, Na and Al are produced and ejected by low- and intermediate-mass stars during the asymptotic giant branch (AGB) phase of stellar evolution (see e.g. \citealt{Bohringer2010, Nomoto2013, Karakas2014} for recent reviews). 

The first observation of a gradient in Fe abundance was made by the \textit{ASCA} observatory in the core of the Centaurus cluster \citep{Allen1994, Fukazawa1994}. Many other groups have since measured an Fe peak, known to be characteristic of a cool-core cluster (e.g. \citealt{Matsushita1997, DeGrandi2001, Gastaldello2002}). As well as an increasing metallicity gradient, cool-core (CC) clusters exhibit decreasing temperature and entropy profiles. This anti-correlation between central entropy and metallicity is well documented (e.g. \citealt{Leccardi2010}). Conversely, non cool-core (NCC) clusters without a decreasing temperature or entropy profile have been observed without a clear increase in central Fe abundance \citep{DeGrandi2001}. 

Outside of the cluster core, X-ray observations have found that there is a remarkable universality and uniformity in the abundance of metals in the ICM (e.g. \citealt{Fujita2008, Leccardi2008, Werner2013, Simionescu2015, Urban2017}. Many groups have looked at the redshift evolution of metallicity in clusters and found that there has been very little evolution in Fe abundance since at least $z = 1$ (e.g. \citealt{Tozzi2003, Baldi2012, McDonald2016, Mantz2017}). By looking at the relative abundance of metals produced in different types of stars, it is possible to determine the relative abundance of Type Ia/Type II SNe and AGB stars, which helps to determine the star formation history of the cluster. Recent studies by \cite{Ettori2015} and \cite{Mernier2017} have found that relative abundance profiles of different metals are also relatively flat.

The observations that metal abundance is relatively constant in cluster outskirts, and has been for the last 6-9~Gyr, has led to the acceptance of the `early enrichment' model of metal abundance. If the majority of iron was recently introduced to a cluster, driven by the release of metals from infalling galaxies via ram-pressure stripping \citep{Domainko2006} together with galactic winds \citep{Kapferer2009}, this would lead to substantial radial gradients in the Fe abundance which would continue beyond the gradient observed in cluster cores, and also introduce azimuthal inhomogeneities (e.g. \citealt{Kapferer2006}). Therefore, to maintain the uniformity in metal abundance observed in cluster outskirts, the bulk of metal enrichment must happen in individual galaxies, before the formation of galaxy clusters, at $z > 2$.

Despite the acceptance of the early enrichment model of metal abundance, the exact processes which contribute to metal enrichment are still unknown. Simulations are powerful tools which can be used to test different stellar enrichment models, and help determine the effects of different astrophysical proccesses on metal enrichment and distribution. A recent review by \cite{Biffi2018b} looks at the progress made over the last few decades in modelling ICM enrichment in hydrodynamical simulations, including looking at the effect of different assumptions/parameters on the resulting metal abundance in simulated clusters. The main factors which impact ICM enrichment all concern the abundance and evolution of different stellar populations. The number of stars produced as a function of mass is set by the initial mass function (IMF), whose shape constrains the relative ratio of massive, short-lived stars (which end as SNII), and low-mass long-lived stars (progenitors of SNIa and AGB stars). The evolution of the different stellar populations depends on the assumed lifetime functions which are typically mass-dependent (e.g. \citealt{Maeder1989, Padovani1993, Chiappini1997}), differing most for the low-mass stars, but can also be metallicity-dependent (\citealt{Portinari1998}, see also \citealt{Romano2005}). Finally, the assumed yield of metals ejected from the different stellar populations is a large source of uncertainty due to the significant variation in published yields (see \citealt{Biffi2018b} and references therein).

Other important mechanisms included in simulations which directly effect the resulting distribution of metals are feedback from stars and AGN. Recent work by e.g. \cite{Sijacki2006}, \cite{Fabjan2010}, and \cite{Biffi2017, Biffi2018a} has shown that AGN feedback is imperative to reproduce observed trends in metallicity, and work by \cite{Dubois2011} has shown that the production of CC clusters also depends on AGN. Although there are still limitations to simulations, many groups are now able to reproduce the observed ratios of CC and NCC clusters, and their different chemical and thermal properties (e.g. \citealt{Rasia2015, Martizzi2016, Biffi2017, Barnes2018, Vogelsberger2018}). The numerical resolution of simulations has also been shown to effect the metal abundance in clusters, with enrichment increasing at higher resolution (e.g. \citealt{Tornatore2007, Vogelsberger2018}) as a result of enhanced star formation at high redshift.

In this paper, we study the spatial distribution and abundance of metals in the C-EAGLE clusters \citep{Barnes2017b, Bahe2017}, and look at how this evolves with redshift. The C-EAGLE simulations are high resolution simulations, incoporating the formation of galaxies down to dwarf scales ($M_{*} \simeq 10^{9}$~M$_{\odot}$) and model the cluster region out to at least 5 virial radii. Being able to analyse the cluster outskirts to such large radii allows us to test in more detail the early enrichment model and see whether the observed homogeneity of metallicity in cluster outskirts is found in our simulated sample. In Section~\ref{Sec:Simulations}, we briefly detail the model used to simulate the C-EAGLE clusters. In Section~\ref{Sec:PresentDay} we look at the present day composition of the ICM through metallicity maps and abundance profiles. We study the redshift evolution of metallicity in Section~\ref{Sec:Redshift}, with a focus on how the core and outskirts of clusters are formed. We summarise our findings in Section~\ref{Sec:Conclusions}.

\section{C-EAGLE sample}\label{Sec:Simulations}

In this paper we use the simulated C-EAGLE clusters \citep{Barnes2017b, Bahe2017}. The C-EAGLE project aimed to investigate the properties of massive galaxy clusters which are typically not included in other simulation suites due to the volume of space needed to form such objects and the associated computational expense. Initially, a cubic periodic volume with side length 3.2~Gpc was simulated using only dark matter (DM) particles \citep{Barnes2017a}, from which the 30 C-EAGLE regions were chosen such that the haloes were evenly spaced in the mass range $14.0 \leq \log_{10}(M_{200}$\footnote{We define the mass of a cluster, $M_{\Delta}$, as the mass within a sphere of radius $r_{\Delta}$ that encloses a density equal to $\Delta$ times the critical density of the Universe, $\rho_{\mathrm{crit}}$. We will typically use overdensity $\Delta = 180, 200$ or 500.}$/\mathrm{M}_{\odot}) \leq 10^{15.4}$ at $z=0$. 

Each cluster was then resimulated using the zoom technique of \cite{Katz1993}, in which the resolution of the area of interest is increased relative to the surroundings, whilst maintaining the correct tidal forces using lower resolution particles outside of the cluster (see also \citealt{Tormen1997}). The simulations assumed a flat, $\Lambda$CDM universe combining the \textit{Planck} 2013 with baryonic acoustic oscillations, \textit{WMAP} polarization and high-multipole experiments \citep{PlanckXVI2014}: $\Omega_{\mathrm{b}} = 0.04825$, $\Omega_{\mathrm{m}} = 0.307$, $\Omega_{\Lambda} =0.693$, $h \equiv H_{0}$/(100 km~s$^{-1}$~Mpc$^{-1}$) = 0.6777, $\sigma_{8} = 0.8288$, $n_{\mathrm{s}} = 0.9611$ and $Y = 0.248$. All DM particles in the high resolution region had mass $m_{\mathrm{DM}} \simeq 9.7\times 10^{6}$~\msol, while the initial gas particle mass was $m_{\mathrm{gas}} \simeq 1.8\times 10^{6}$~\msol. The gravitational softening length was set to 2.66~kpc (co-moving) for $z > 2.8$, and fixed in physical coordinates to 0.70~kpc for $z < 2.8$. The minimum SPH smoothing length was set to 0.1 times the gravitational softening at all times. 
	
During the simulations, an on-the-fly friends-of-friends (FOF) algorithm was used to link all particles within a dimensionless linking length, $b=0.2$, which can lead to the inclusion of non-gravitationally bound particles \citep{Davis1985}. Instead, \textsc{subfind} \citep{Springel2001,Dolag2009} was used to refine the FOF groups by identifying the most bound particle, and removing all particles which are not gravitationally bound to a distinct sub-halo.

\subsection{The EAGLE model}

The C-EAGLE clusters were simulated using the EAGLE code \citep{Schaye2015}, which is a heavily modified version of \textsc{gadget-3} (last described as \textsc{gadget-2} by \citealt{Springel2005b}) and uses a subgrid model based on that developed for the OWLS project \citep{Schaye2010}. The smoothed particle hydrodynamics (SPH) algorithms, collectively referred to as \textsc{anarchy}, are described in detail by \citeauthor{Schaye2015} (\citeyear{Schaye2015}, Appendix A) and \cite{Schaller2015}. \textsc{anarchy} uses the pressure-entropy formalism of \cite{Hopkins2013}, an artificial viscosity switch \citep{Cullen2010}, an artificial conductivity switch \cite{Price2008} and the time-step limiter of \cite{Durier2012}. The SPH smoothing lengths were calculated using the $C^{2}$ kernel \citep{Wendland1995} and 58 nearest-neighbour particles. The EAGLE subgrid model is described in detail in \cite{Schaye2015}. Here, we briefly describe the aspects of the subgrid model which most affect the metal production and distribution.

Radiative cooling is calculated on an element-by-element basis following the model of \cite{Wiersma2009a}. Cooling rates for 11 elements tracked throughout the simulation are tabulated as a function of density, temperature and redshift using \textsc{cloudy} version 07.02 \citep{Ferland1998}, assuming ionization equilibrium, the presence of the cosmic microwave background, and an evolving UV/X-ray background from galaxies and quasars (implemented using the model of \citealt{Haardt2001}). The tracked elements are: H, He, C, N, O, Ne, Mg, Si, S, Ca and Fe.

Stars are formed at a pressure-dependent rate from gas particles which obey an equation of state $P \propto \rho^{4/3}$, where $P$ is gas pressure and $\rho$ is gas density, according to the stellar formation model of \cite{Schaye2008}. The metallicity-dependent density threshold of \cite{Schaye2004} is used, $n_{\mathrm{H}} > 0.1~\mathrm{cm}^{-1}(Z/0.002)^{-0.64}$, such that above this density a cold gas phase is expected to form stars. Star formation is implemented stochastically and assumes a \cite{Chabrier2003} initial mass function (IMF).

Stellar evolution and enrichment are implemented using the model of \cite{Wiersma2009b}. At each time-step the metallicity-dependent lifetimes of \cite{Portinari1998} are used to determine which stellar particles have reached the end of the main sequence phase. The model assumes that mass is lost from these stars via winds from asymptotic giant branch (AGB) stars, Type Ia and Type II supernovae. The nucleosynthetic yields of \cite{Marigo2001} and \cite{Portinari1998} are used for AGB stars and Type II supernovae, while the yields from the W7 model of \cite{Thielemann2003} are used for Type Ia supernovae. The mass lost by a star particle is distributed to the 48 nearest neighbour particles determined using the SPH kernel, assuming that all SPH particles have the same mass to stop runaway enrichment of a single gas particle.

The EAGLE model uses the stochastic thermal stellar feedback model of \cite{Dalla2012}, in which the probability of a particle being heated is related to the fraction of the total energy per core collapse supernova which is injected on average per unit stellar mass, $f_{\mathrm{th}}$. Heating events happen once in the lifetime of the stellar particle when it has reached $3\times 10^{7}$ yr, corresponding to the maximum lifetime of a star which explodes as a Type II supernova. The temperature jump of a heated particle is $\Delta T = 10^{7.5}$~K.

Black holes are seeded at the centres of FOF groups with mass greater than $10^{10}\text{ M}_{\odot}/h$ by converting the most dense gas particle to a BH seed particle, $m_{\text{BH}} = 10^5\text{ M}_{\odot}/h$. BHs can then grow via gas accretion or mergers. BH feedback is implemented stochastically such that once the BH energy is above a critical energy, the BH has probability $P$ of heating a single particle by an amount $\Delta T$. The C-EAGLE project uses the AGNdT9 model, for which $\Delta T = 10^9$~K, as it has been shown to produce a good match to the observed gas mass fractions of low mass groups \citep{Schaye2015}. If there is still energy left above the feedback threshold after a heating event, the timestep of the BH is shortened so that more energy can be released.

\section{Metals in the present-day ICM}\label{Sec:PresentDay}

\begin{figure*}
	\includegraphics[width=\textwidth, trim=3cm 2.5cm 2.5cm 2.5cm, keepaspectratio=True]{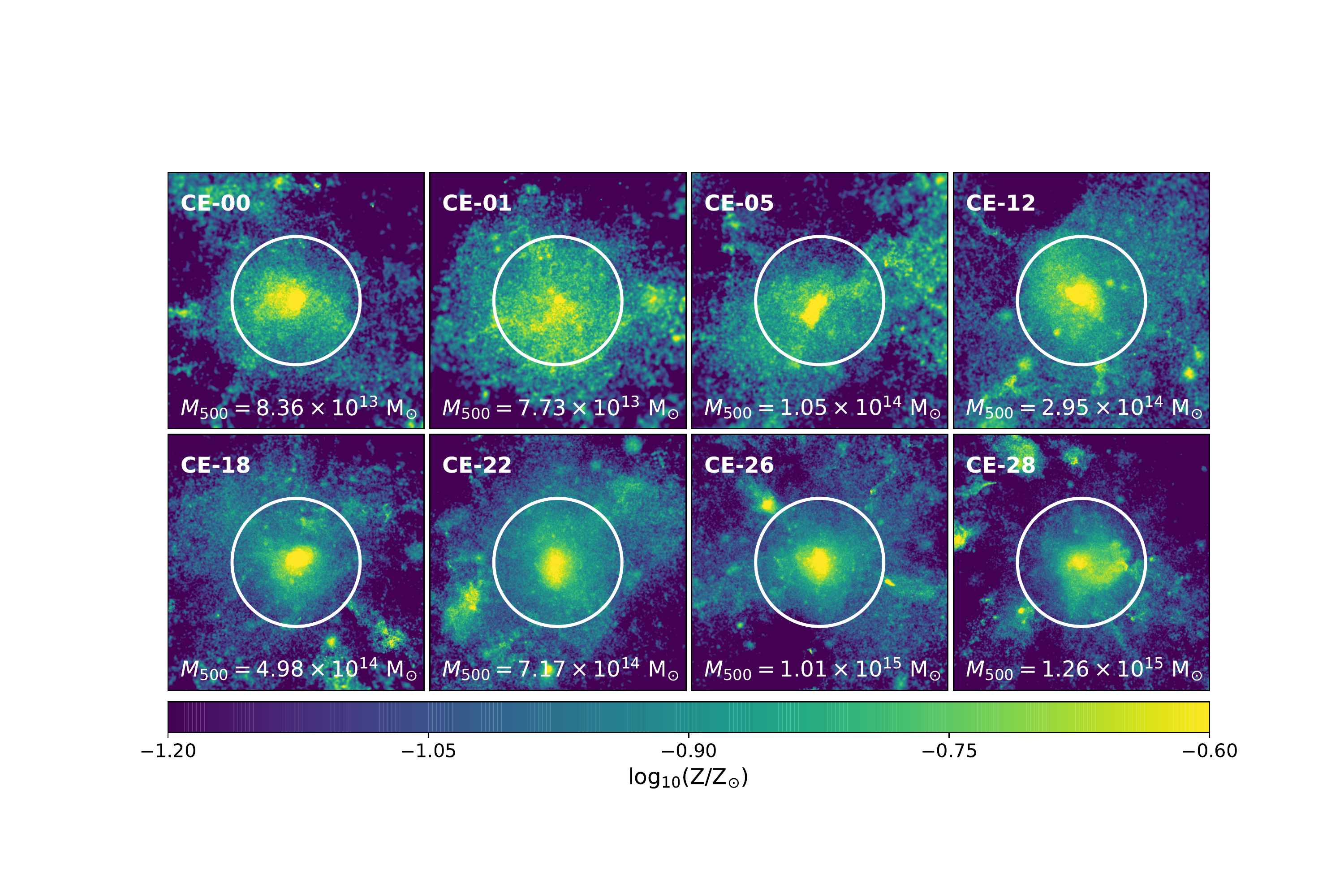}
	\caption{Total metallicity maps for a subset of the C-EAGLE clusters at $z=0$. The white circle shows the position of $r_{500}$, while the $M_{500}$ mass of each cluster is given in each subplot. All maps have an extent of 4~$r_{500}$ in all directions and are projected down the $z$-axis. Gas particles with $T > 10^{5}$~K and $n < 0.1$~cm$^{-3}$ in the region of interest are included in the maps. The metallicity increases towards the centre of all the clusters reaching a peak of $\sim 0.25$ Z$_{\odot}$, decreasing to $\sim 0.1$ Z$_{\odot}$ at $r_{500}$.}
	\label{fig:map}
\end{figure*}

In this section we examine the present-day composition of the ICM, focussing on the total metallicity in the C-EAGLE clusters as well as the abundance of iron, silicon and oxygen. First, we present total metallicity maps for a representative sample of clusters at $z=0$, looking at the distribution of metals throughout individual clusters and considering any trend of metallicity with mass or cluster temperature. We present mass-weighted abundances, given by
\begin{equation}\label{eq:Zmw}
Z_{\mathrm{mw}} = \dfrac{\sum_{i}m_{i}Z_{i}}{\sum_{i}m_{i}},
\end{equation}
where the summation is over the gas particles, $Z$ is the chosen metal (Z, Fe, Si or O), and $m$ is the particle mass. Then we move onto the median metallicity and relative abundance profiles, looking for any radial trends in the range $0.01 < r/r_{500} < 7.0$, where 7~$r_{500}$ is within 5~$r_{200}$ for all of the clusters analysed. All profiles that we present are three-dimensional, mass-weighted quantities that are normalised by the solar abundances of \cite{Anders1989}. Specifically, we use Z$_{\odot} = 1.941\times 10^{-2}$, Fe$_{\odot} = 1.83\times 10^{-3}$, Si$_{\odot} = 6.98\times 10^{-4}$ and O$_{\odot} = 9.54\times 10^{-3}$. Any observational data we present in this paper has been renormalised to the values of \cite{Anders1989}, and we transform between different radii using R500:R200:R180 = 0.62:0.95:1, which corresponds to the median value of each ratio for the C-EAGLE sample.

Throughout this paper we adopt the same naming scheme as in \cite{Bahe2017} and \cite{Barnes2017b}, and refer to individual clusters as CE-X where X is the halo number from the C-EAGLE analysis. We only use 29 of the C-EAGLE clusters, instead of the full 30, due to CE-27 undergoing a large AGN explosion at very high redshift which removes the majority of the gas from the cluster, resulting in an object that would not be included in any observational sample.

From \cite{Barnes2017b}, none of the C-EAGLE clusters are defined as cool-core using the central entropy. This was based on the common definition that a cluster is CC if the cooling time of their central region is short compared to the age of the Universe. When looking at the metallicity content of clusters it is common to separate CC and NCC clusters and analyse the two groups separately due to observations which suggest that the metal abundance depends on the strength of the CC (e.g. \citealt{DeGrandi2001, DeGrandi2004}). This limits our ability to be able to reproduce the observed trends as the C-EAGLE simulations do not accurately reproduce the observed CC cluster fraction.

Another way to separate clusters is by dynamical state. Using the method of e.g. \cite{Barnes2017b, Barnes2018, Barnes2019}, a cluster is defined as relaxed if
\begin{equation}
E_{\mathrm{kin},500} < 0.1~E_{\mathrm{therm},500},
\end{equation}
where $E_{\mathrm{kin},500}$ is the sum of kinetic energy of the gas particles within $r_{500}$ after the bulk motion of the cluster has been removed, and $E_{\mathrm{therm},500}$ is the sum of the thermal energy within the same radius. At $z=0$, 9 of 29 clusters are defined as relaxed.

Using mock spectroscopic (spec) data, generated using the method of \cite{LeBrun2014}, \cite{Barnes2017b} found that the mass-weighted iron abundance within $r_{500,\mathrm{spec}}$ of the C-EAGLE clusters, plotted as a function of spectroscopic temperature, was in reasonable agreement with the consolidated observational data of \cite{Yates2017}. As we use the true simulation data in our analyses and do not attempt to produce mock spectroscopic data, we compare the true and spec data in Appendix~\ref{ap:comp} to show that the results in this paper would not change if we had attempted to produce data, which more closely resembles that used in observational analyses.

\subsection{Metallicity maps}\label{Sec:Maps}

The metallicity maps for a subset of the C-EAGLE clusters can be seen in Fig.~\ref{fig:map}. Here we present the total metallicity in each cluster, normalised by the solar metallicity values of \cite{Anders1989}. Each map has side length $4~r_{500}$ and are projected $4~r_{500}$ along the $z$-axis. All gas particles with $T > 10^{5}$~K and $n < 0.1$~cm$^{-3}$ within the region of interest around the clusters are included in the maps. The white ring corresponds to $r_{500}$, and the halo name and mass of each cluster in solar mass units is given within each of the subplots. All of the maps were created with the same metallicity limits and share the colour bar shown at the bottom of the figure. We include a mixture of relaxed (CE-00, 12, 22, 26) and unrelaxed clusters (CE-01, 05, 18, 28) which span the mass range of the full C-EAGLE sample.

From Fig.~\ref{fig:map}, all of the clusters show the same trend of increasing metallicity with decreasing radius towards the centre of the cluster. The presence of a central metallicity peak has been widely observed in CC clusters, while NCC clusters are found to have flatter metallicity distributions \citep{DeGrandi2001, DeGrandi2004, dePlaa2006, Leccardi2008, Sanderson2009, Lovisari2011, Mernier2015, Mernier2017}. The metallicity rich cores peak at around 3~Z$_{\odot}$ (although our maps are clipped at 0.25~Z$_{\odot}$ for clarity) and extend to 0.2~$r_{500}$, at which point the value of $Z$ drops to around 0.15~Z$_{\odot}$ at $r_{500}$, as expected from observations. Although the overall trends are the same across the cluster sample, there are still small scale differences in the metallicity distributions between the clusters showing that the ICM is not fully homogeneous.

\begin{figure}
	\includegraphics[width=\columnwidth,keepaspectratio=True,trim=0cm 1cm 0cm 0cm]{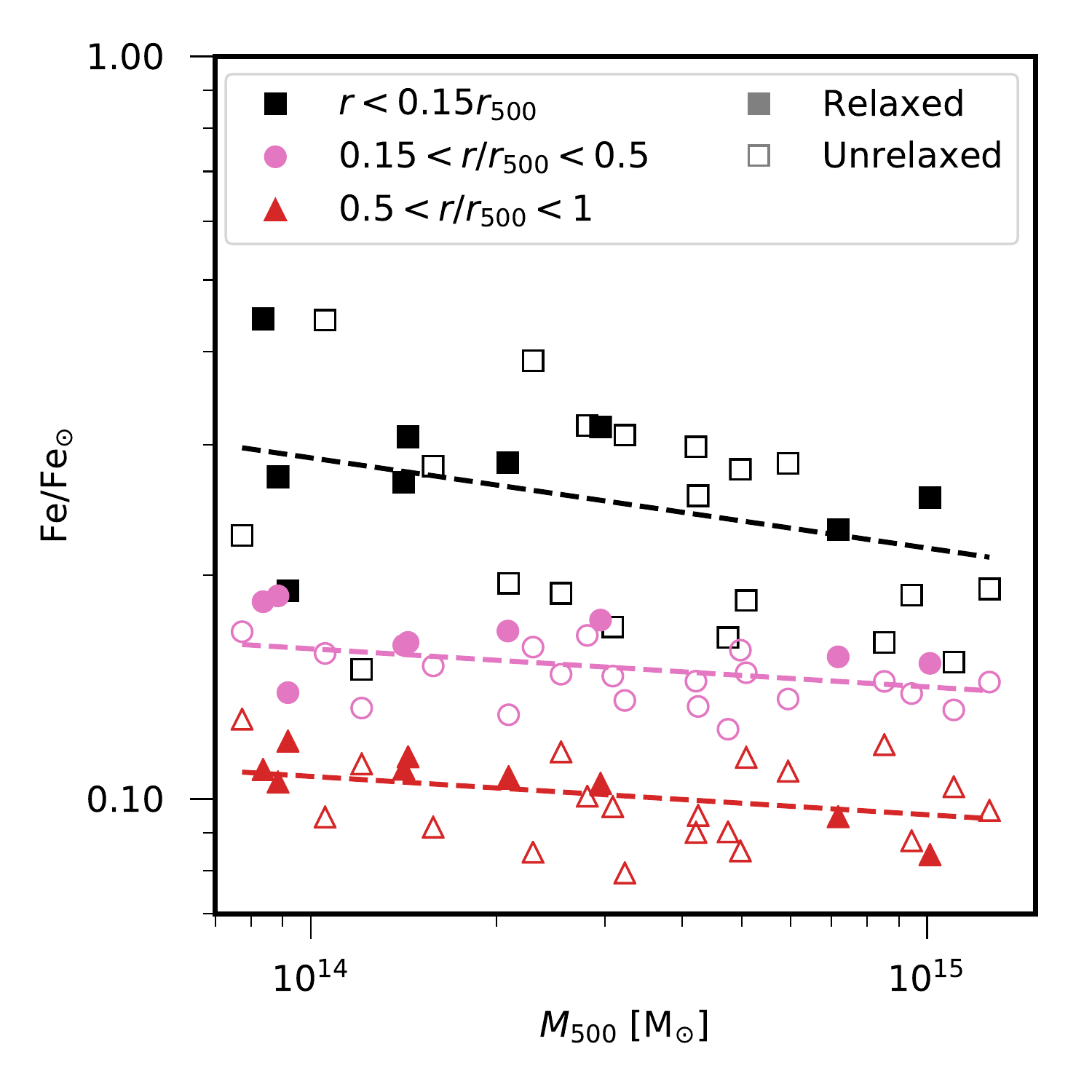}
	\caption{Fe abundance within different 3D apertures, plotted against $M_{500}$, for the 29 C-EAGLE clusters used in this analysis. The black squares represent the metallicity within $0.15~r_{500}$, pink circles are $0.15 < r/r_{500} < 0.5$, and red triangles are $0.5 < r/r_{500} < 1.0$. The best-fit line to the different datasets assuming the model given by Eq.~\ref{eq:mass} are shown as dashed lines in the same colour as the individual data points.}
	\label{fig:Fe_M500}
\end{figure}

Observational studies looking at the metallicity outside of the core have found that there is very little azimuthal dependence of metallicity (e.g. \citealt{Werner2013, Simionescu2015}), as was seen in Fig.~\ref{fig:map}. From the C-EAGLE maps, there is typically an azimuthal dependence of the extended high metallicity core which is not spherical in the majority of the clusters. This suggests that dynamics in the core effects the distribution of metals, although the level of metal enrichment is believed to have been set at a much higher redshift, $z > 2$, according to the early enrichment model. At $r > r_{500}$, it can also be seen that the metallicity follows the large-scale structure, with objects external to the clusters being metal rich.

\subsection{Metallicity - mass relation}

Looking at the cluster maps, Fig.~\ref{fig:map}, there is no obvious trend of metallicity with mass (and therefore, temperature). We look at this in more detail in Fig.~\ref{fig:Fe_M500} which shows the Fe abundance within three different apertures for all of the C-EAGLE clusters. The cluster core ($r < 0.15~r_{500}$) is represented by black squares, with the pink circles and red triangles showing the metallicity in increasingly larger apertures, $0.15 < r/r_{500} < 0.5$ and $0.5 < r/r_{500} < 1.0$ respectively. (Un)relaxed clusters are presented as (open)filled data points. The points in each aperture is fit with a simple power-law model,
\begin{equation}\label{eq:mass}
\mathrm{Fe} = \mathrm{Fe}_{0}\left(\dfrac{M_{500}}{3\times 10^{14}~\mathrm{M}_{\odot}}\right)^{\beta}, 
\end{equation}
where Fe$_{0}$ is a normalisation constant and $\beta$ encompasses the correlation between metallicity and mass. The pivot point of $3\times 10^{14}$~M$_{\odot}$ is used as it is the median of the C-EAGLE sample. The best-fit relations are shown as a dashed line in the same colour as the corresponding aperture.

There is a mild anti-correlation between metallicity and mass in all three of the chosen apertures. There is most evidence for a mass dependence of Fe in the core of the clusters, which has also been seen by \cite{Truong2019} using a different simulated cluster sample. Using a homogenised observational dataset, \cite{Yates2017} also find evidence for a mild anti-correlation between Fe and temperature, often used as a proxy for mass, which is present regardless of whether the core is excised. Recently, \cite{Lovisari2019} have suggested that the anti-correlation between metallicity and mass/temperature is driven by relaxed clusters which have not recently undergone a merger and thus the metal rich core is untouched. They argue that in unrelaxed clusters the metallicity abundance is mixed due to larger scale motions as a result of mergers. However, in the C-EAGLE clusters, we find that the clusters defined as unrelaxed (open points in Fig.~\ref{fig:Fe_M500}) also show a mild anti-correlation between central metallicity and mass.

\begin{table}
	\begin{center}
		\caption{The best-fit parameters for the mass-metallicity relation, Eq.~\ref{eq:mass}, in three different apertures. The errors are found by bootstrapping the data 1000 times and taking the 16th and 84th percentiles of the best-fit parameters of the bootstrapped data for the lower and upper errors respectively.}
		\label{tab:Fe_M500_fit}
		\begin{tabular}{l|cc}
			& Fe$_{0}$ & $\beta$\\
			\hline
			$r < 0.15~r_{500}$ & $0.25\pm 0.01$ & $-0.15\pm0.07$\\
			$0.15 < r/r_{500} < 0.5$ & $0.15\pm 0.01$ & $-0.06\pm0.02$\\
			$r > 0.5~r_{500}$ & $0.10\pm0.01$ & $-0.06\pm0.03$\\
		\end{tabular}
	\end{center}
\end{table}

The best fit parameters of Eq.~\ref{eq:mass} and their errors, obtained from bootstrapping the data 1000 times, are shown in Table~\ref{tab:Fe_M500_fit}. As expected, the normalisation Fe$_{0}$ is larger in the smallest aperture where the metallicity values peak, and decreases as you get further from the cluster centre. There is also increased uncertainty in the $\beta$ value for the central region due to the increased scatter in metallicity in the core.

\subsection{Median iron \& total metallicity profiles}

\begin{figure}
	\includegraphics[width=\columnwidth,keepaspectratio=True, trim=0cm 1cm 0cm 0cm]{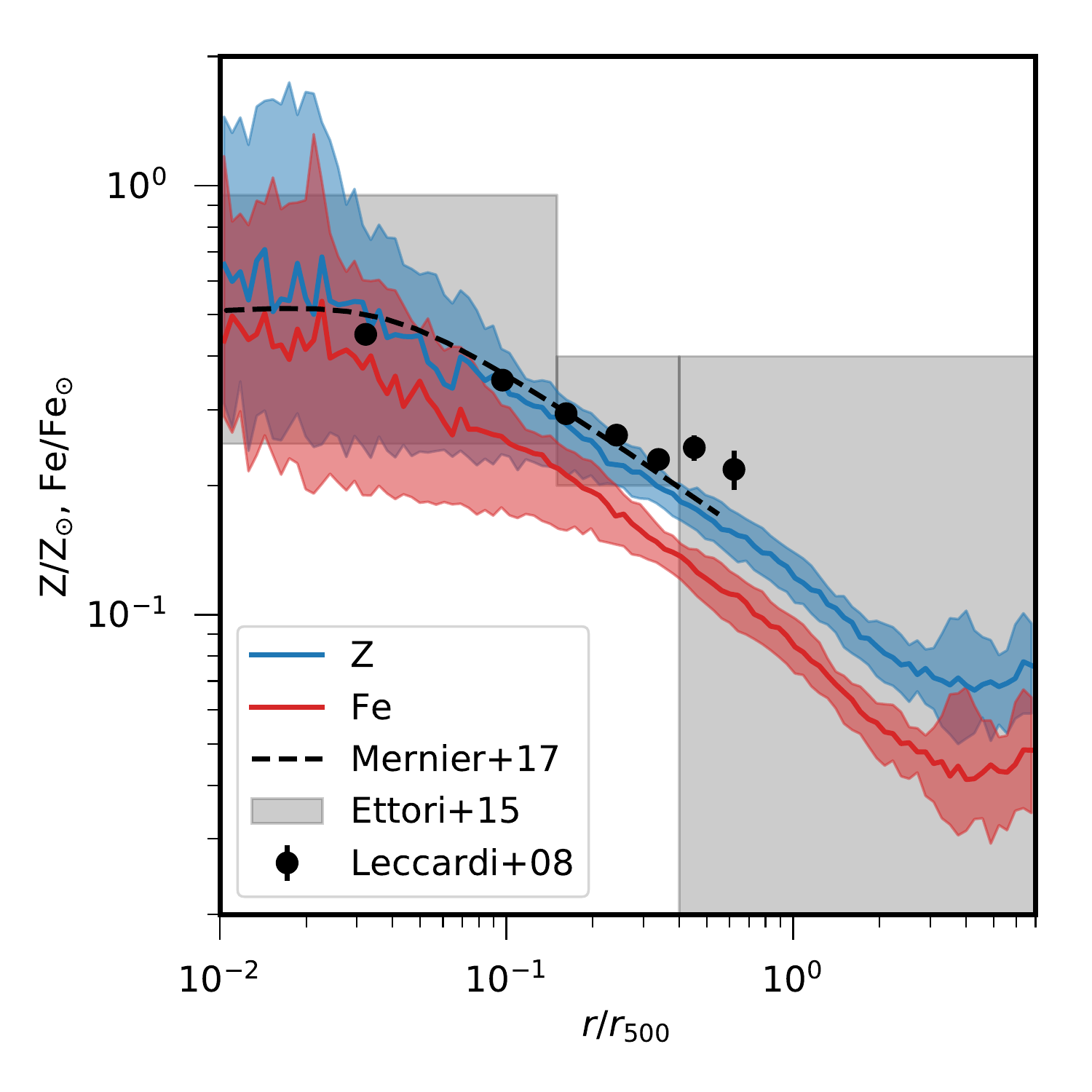}
	\caption{The median mass-weighted ICM metallicity (blue) and iron abundance (red) profiles at $z=0$. The coloured shaded regions show the 1$\sigma$ scatter in the data across the C-EAGLE clusters. For comparison we also include the empirical fit to Fe obtained by \citeauthor{Mernier2017} (\citeyear{Mernier2017}, dashed black line), the measured Fe data points obtained by \citet{Leccardi2008} and the scatter in Fe from \citet{Ettori2015}.}
	\label{fig:Z_Fe_comparison}
\end{figure}

We present the median total metallicity (blue) and Fe abundance (red) profiles at $z=0$ in Fig.~\ref{fig:Z_Fe_comparison}. The coloured shaded regions show the $1~\sigma$ scatter in the profiles, taken between the 16th and 84th percentiles. The black dashed line is the empirical fit to the Fe abundance of the CHEERS sample \citep{dePlaa2017, Mernier2016} obtained by \cite{Mernier2017},
\begin{equation}
\dfrac{\mathrm{Fe}(r)}{\mathrm{Fe}_{\odot}} = 0.21(r+0.021)^{-0.48}-6.54\exp\left(-\dfrac{(r+0.0816)^{2}}{0.0027}\right),
\end{equation}
where $r$ is measured in units of $r_{500}$, and we have re-normalised to the solar abundance of \cite{Anders1989} from the proto-solar value of \cite{Lodders2009}. The black circles are observational data points from \cite{Leccardi2008}, and the grey shaded regions show the scatter in Fe abundance obtained by \cite{Ettori2015}.

\begin{figure}
	\includegraphics[width=\columnwidth,keepaspectratio=True, trim=0cm 1cm 0cm 0cm]{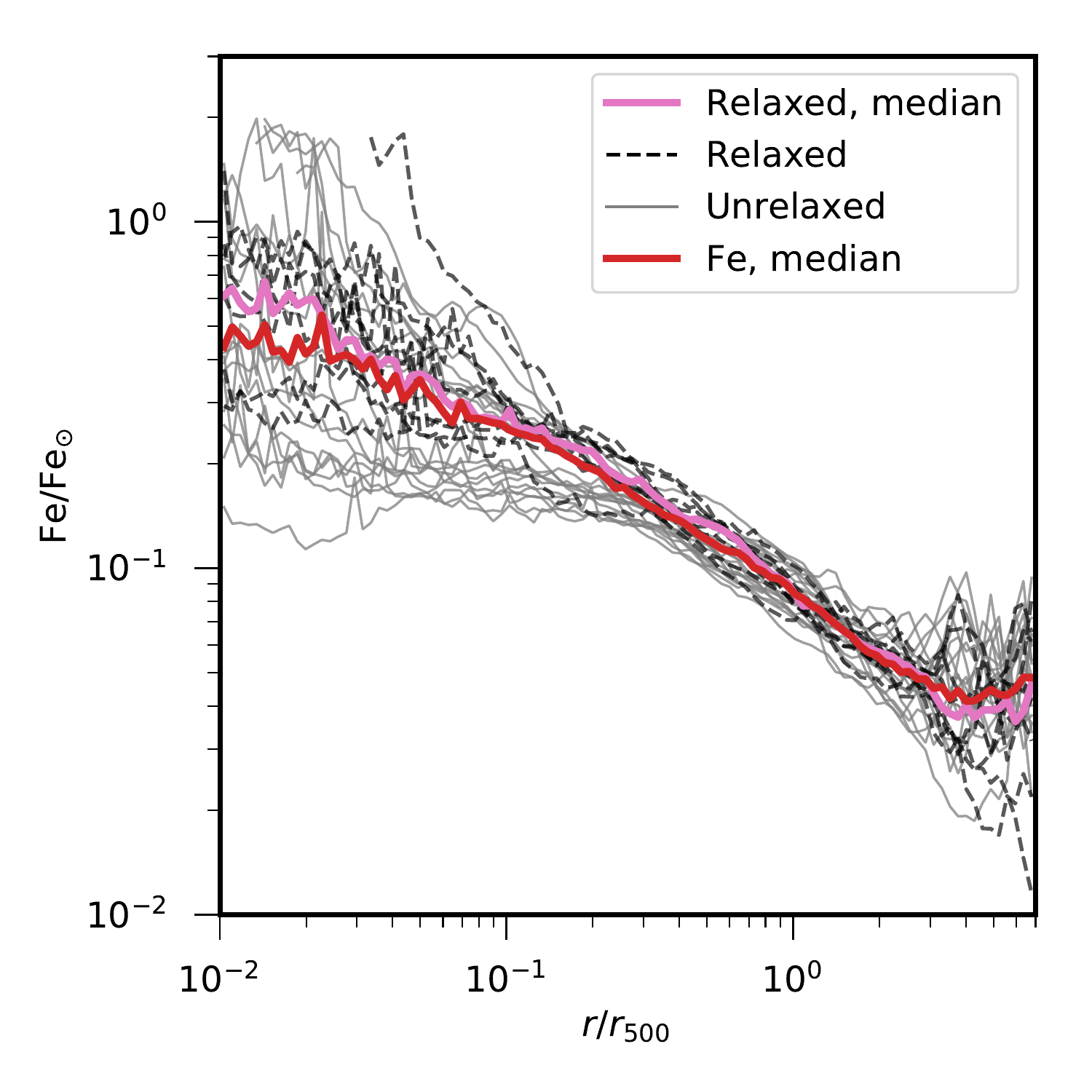}
	\caption{The median mass-weighted iron profile is shown in red, with the median of the relaxed cluster subset shown in pink. Individual profiles of unrelaxed clusters are presented in grey, with relaxed clusters shown as black dashed lines. Below $r < 0.2~r_{500}$ there appears to be two distinct classes of cluster in which the metallicity continues to increase or it plateaus. The relaxed clusters all continue to increase, but so do a number of the unrelaxed clusters.}
	\label{fig:Fe_profile}
\end{figure}

Comparing the Fe profile to the observations, the C-EAGLE cluster simulations do not produce enough iron by around a factor of 1.5. This is consistent with what was found by \cite{Barnes2017b} when using emission-weighted median Fe profiles of the C-EAGLE clusters\footnote{Here we are comparing to fig.~15 in \cite{Barnes2017b} which appears to show good agreement between the emission-weighted, median Fe profile for the C-EAGLE clusters and observed profiles. However, it was found that there was an issue with the original analysis which introduces a radial shift that causes the profiles to artificially line up. When correcting for this radial shift, the profiles show the same offset as seen in Fig.~\ref{fig:Z_Fe_comparison}.}, as opposed to the mass-weighted profiles used in this analysis. We have investigated the effect of using mock spectroscopic data to more closely reproduce X-ray observational anlyses and found that there is very little difference between the true and spectroscopic abundance profiles (see Appendix~\ref{ap:comp} for more details). Therefore, this difference in normalisation is not likely to be caused by using the mass-weighted profiles. The EAGLE model was not calibrated to reproduce any metal abundance, so given the uncertainty in metal yields, the inital mass function and supernova rates, it is not unsurprising that the simulation has Fe abundance that is too low. From \cite{Barnes2017b}, selecting the relaxed subset of C-EAGLE clusters does increase the normalisation of the Fe abundance profile, but still not to the same level as the observations.

\begin{figure}
	\includegraphics[width=\columnwidth,keepaspectratio=True,trim=0cm 1cm 0cm 0cm]{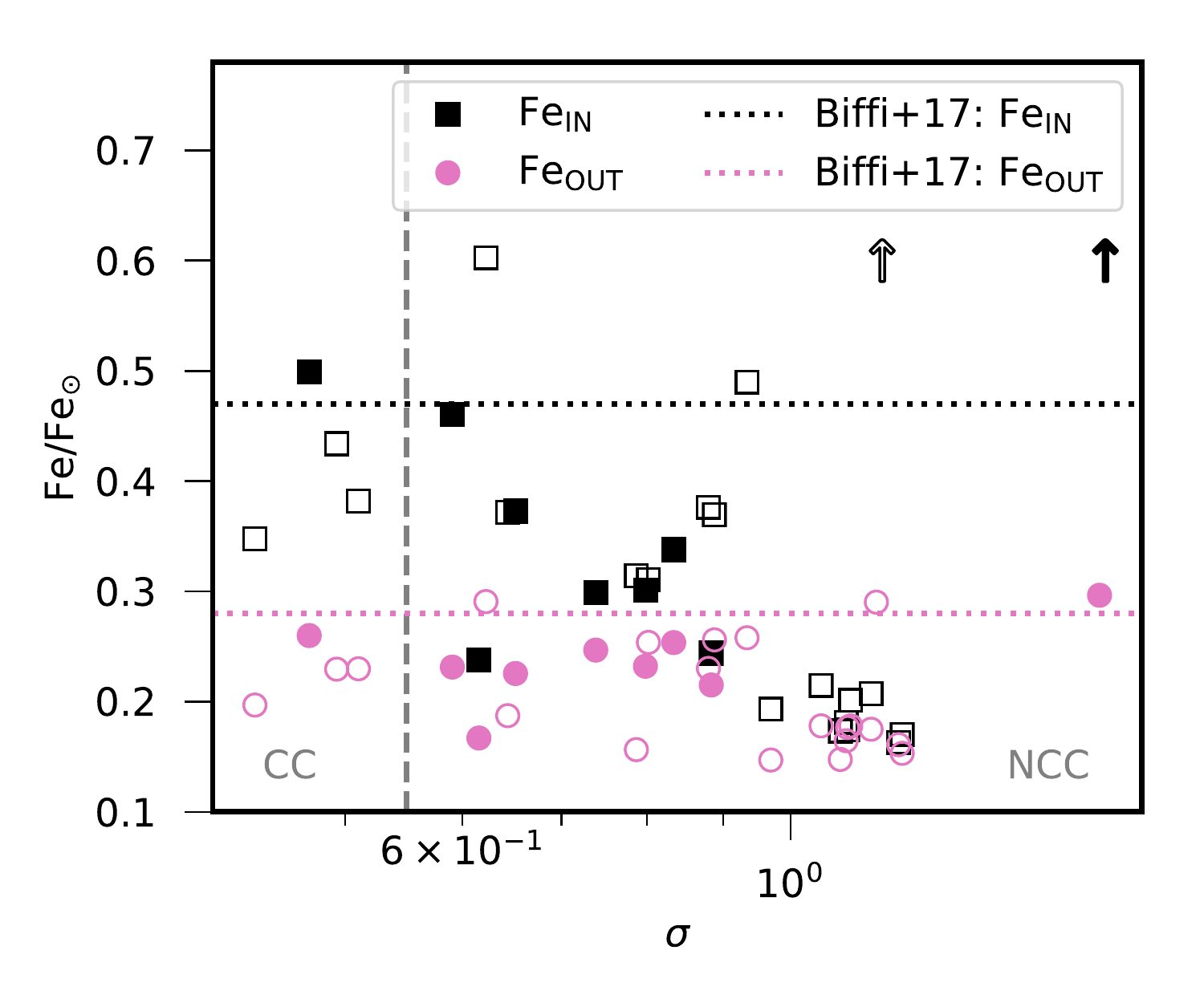}
	\caption{The mass-weighted Fe abundance within two different apertures plotted against the pseudo-entropy $\sigma$ (Eq.~\ref{eq:pseudo-entropy}, see also e.g. \citet{Rasia2015}) for each cluster. The black squares correspond to the aperture $r < 0.08~r_{500}$, while the pink circles are $0.08 < r/r_{500} < 0.32$. The two black arrows correspond to two clusters which have Fe$_{\mathrm{IN}} > 0.8$. The vertical dashed grey line gives the value below which clusters are defined as having a cool-core. The horizontal dotted lines give the median values of Fe in the different apertures found by \citet{Biffi2017}.}
	\label{fig:Fe_sigma}
\end{figure}

Looking at other simulations, \cite{Vogelsberger2018} find a similar, if slightly smaller, offset between observations and the median Fe profiles for the IllustrisTNG cluster sample. They present results for two different mass resolutions which suggests that the Fe abundance decreases at lower resolution (comparison of TNG100-1 and TNG300-1, where the TNG300-1 data has been scaled to the higher resolution but smaller volume TNG100-1 simulation by a correction factor of 1.6). This is driven by the resolution dependent star formation, as less stars results in less metals. However, \cite{Biffi2017} find good agreement with the observations of \cite{Leccardi2008} despite using a lower mass resolution simulation than both IllustrisTNG and C-EAGLE. All three simulations use a \cite{Chabrier2003} IMF, but have very different yields and stellar lifetimes implemented, as well as using different models for stellar evolution and chemical enrichment. Although \cite{Tornatore2007} found that changing the IMF had the biggest impact on Fe abundance, comparison of these three simulations shows that a large number of assumptions and physical processes affect the abundance and distribution of metals in the intracluster medium, making it increasingly difficult to determine what controls the normalisation of the abundance profiles.

\begin{figure*}
	\includegraphics[width=\textwidth,keepaspectratio=True,trim=0cm 1cm 0cm 0cm]{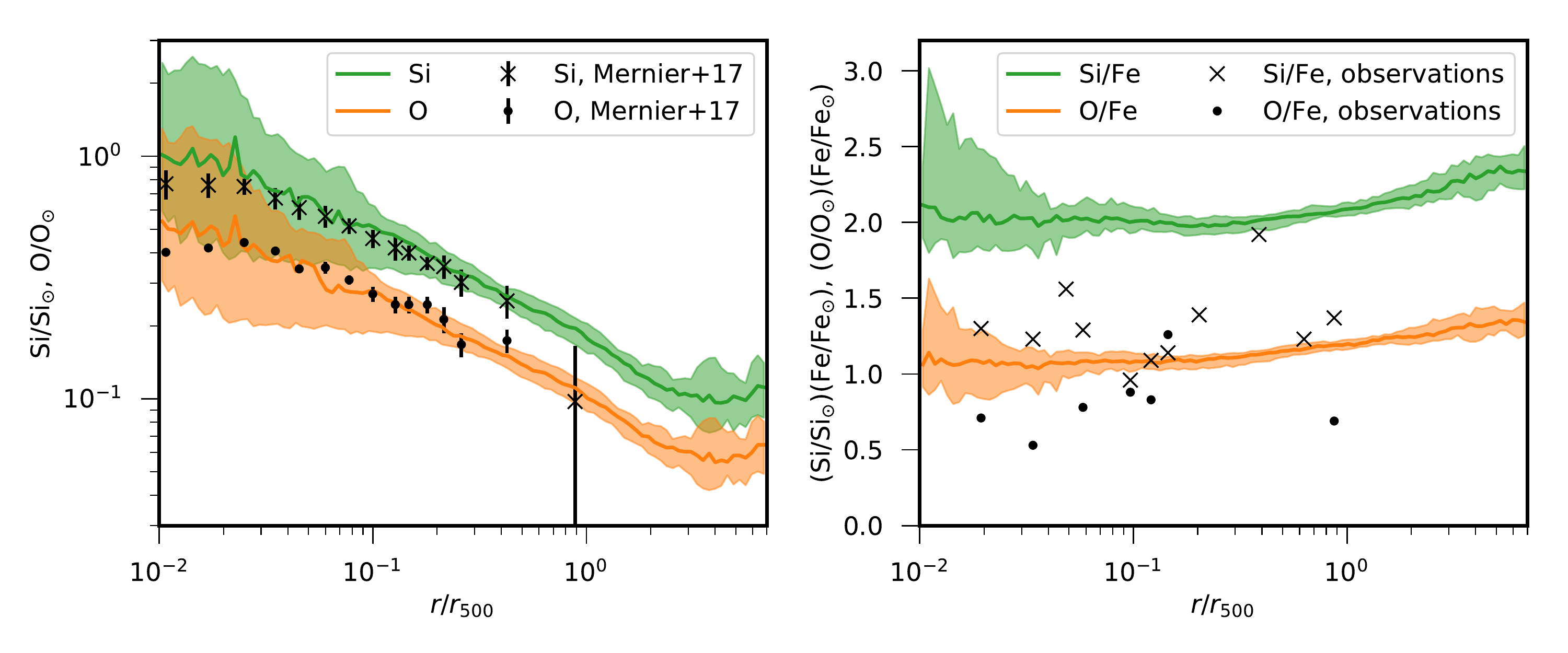}
	\caption{Left: The silicon and oxygen mass-weighted median abundance profiles, presented in green and orange, respectively. The coloured bands show the 1$\sigma$ scatter in the respective profiles. The black crosses and circles show the average metallicity for Si and O, respectively, found by \citet{Mernier2017}. Right: The median abundance ratios of silicon and oxygen with respect to iron. The scatter in the individual abundance ratios is again shown as a coloured band around each profile. The black crosses and circles show the measurements of Si/Fe and O/Fe, respectively, from observations of AWM7 \citep{Sato2008}, Centaurus \citep{Sakuma2011} and Coma (only Si/Fe, \citealt{Matsushita2013}). The offset between the observed abundance ratios and those obtained from the C-EAGLE clusters is a result of the Fe abundance in C-EAGLE being too low.}
	\label{fig:Si_O_comparison}
\end{figure*}

Although the normalisation of the median Fe profiles is too low, the shape of the profile is in very good agreement with the observations of \cite{Leccardi2008} and \cite{Mernier2017}, out to at least $0.3~r_{500}$. At larger radii, the Fe abundance in the C-EAGLE clusters continues to decline linearly until around $3~r_{500}$, where the profile begins to flatten. There is some observational evidence that metallicity should be relatively constant at $r > 0.4~r_{500}$, see \cite{Mernier2018} fig.~3 and references therein. However, measurements of metallicity at radii beyond half the virial radius of clusters are difficult to obtain due to the low X-ray surface brightness in cluster outskirts. It is hoped that the next generation of X-ray telescopes will help to characterise the metallicity distribution of clusters at large radii.

From Fig.~\ref{fig:Z_Fe_comparison}, the scatter in the abundance profiles is significantly larger in the core, $r < 0.2~r_{500}$, than at any other radius. In order to look at this in more detail, we present the individual Fe abundance profiles of the 29 C-EAGLE clusters in Fig.~\ref{fig:Fe_profile}. The red line again shows the median mass-weighted Fe abundance profile, while the pink line is the median Fe profile of the relaxed subset of clusters. The individual profiles of the relaxed sample are given by the black dashed lines, while the grey solid lines are the unrelaxed clusters. Within the core of the clusters the scatter in the profiles shows that there are two distinct groups; the metallicity either continues to increase towards the centre of the cluster, or the metallicity is constant at small radii. This bimodal distribution is typically attributed to the difference between cool-core (CC) and non-cool-core (NCC) clusters which have increasing and constant metallicity respctively. However, from \cite{Barnes2017b}, all of the C-EAGLE clusters are NCC when using central entropy. 

Instead, we choose to separate the C-EAGLE clusters by their dynamical state as recent work by \cite{Lovisari2019} suggests that the observed bimodality in metallicity profiles could be driven by increased central abundance of relaxed clusters. Looking at the individual profiles, the relaxed subset (dashed lines) typically do all have increasing metallicity at decreasing radii. The median of the relaxed clusters does show good agreement with the full sample at large radii, and shows the expected deviation at low radii. However, there is a lot of scatter in the clusters which are not defined as relaxed, with a number of them also having increasing metallicity profiles. In general, it does not seem that relaxation is a good proxy for CC/NCC, especially at high redshift where the fraction of CC clusters is expected to increase while the percentage of relaxed clusters decreases (e.g. \citealt{Barnes2018}).

As metallicity is expected to be anti-correlated with cluster entropy, another way to define clusters as CC is to use the pseudo-entropy,
\begin{equation}\label{eq:pseudo-entropy}
\sigma = \left(\dfrac{T_{\mathrm{sl,IN}}}{T_{\mathrm{sl,OUT}}}\right)\left(\dfrac{\mathrm{EM}_{\mathrm{IN}}}{\mathrm{EM}_{\mathrm{OUT}}}\right)^{-1/3}
\end{equation}
where $T_{\mathrm{sl}}$ is the spectroscopic-like temperature \citep{Mazzotta2004}, and EM is the emission-measure metallicity, found by including the gas density in Eq.~\ref{eq:Zmw},
\begin{equation}
EM = \dfrac{\sum_{i}m_{i}\rho_{i}Z_{i}}{\sum_{i}m_{i}\rho_{i}},
\end{equation}
Following \cite{Rasia2015}, the `IN' and `OUT' regions are $ r < 0.05 r_{180}$ and $0.05 < r/r_{180} < 0.2$ respectively, which converting to $r_{500}$ gives $r < 0.08~r_{500}$ and $0.08 < r/r_{500} < 0.32$. Clusters are then defined as CC if $\sigma < 0.55$. Using this definition there are four clusters within the C-EAGLE sample that can be thought of as CC, whereas there are no CC clusters when using central entropy. Compared to the simulations of \cite{Biffi2017}, the C-EAGLE sample has fewer clusters with low values of $\sigma$, but relatively similar maximum value of $\sigma$ at $z=0$. 

We plot the Fe abundance in the IN and OUT regions against $\sigma$ in Fig.~\ref{fig:Fe_sigma} as the black squares and pink circles respectively. Two of the clusters have very high Fe$_{\mathrm{IN}}$ values, and are represented by black arrows situated at the correct $\sigma$ value but at a lower value of Fe$_{\mathrm{IN}}$. Looking in more detail at the metallicity/gas properties of these two clusters, they both have metallicity peaks at small radius which look like they could be the result of a merging object which is just reaching the centre of the cluster. There is no corresponding peak in the density profiles, suggesting that these mergers happened at a much earlier time but the metal rich gas from the centre of the other object has only just reached the core of the cluster. We present (un)relaxed objects as (open)filled points.

Focussing on the IN aperture, there is a general trend of decreasing metallicity with increasing entropy, as expected from the observed anti-correlation. The general trend of decreasing Fe$_{\mathrm{IN}}$ with increasing $\sigma$ is also seen by \cite{Biffi2017}. For comparison, we include the median values of Fe$_{\mathrm{IN}}$ and Fe$_{\mathrm{OUT}}$ obtained by \cite{Biffi2017} as the horizontal dotted lines (red and black respectively). The C-EAGLE median values are noticeably lower in both apertures, again highlighting how the C-EAGLE clusters have an Fe abundance that is too low compared to both other simulations and observations (Fig.~\ref{fig:Z_Fe_comparison}).

In the OUT aperture there is essentially no trend of metallicity with $\sigma$, as expected from \cite{Biffi2017}. This can also be seen in the Fe abundance profiles, Fig.~\ref{fig:Fe_profile}, which shows that the scatter in the individual profiles decreases beyond $0.1~r_{500}$, until around $2~r_{500}$. The two clusters with very high Fe$_{\mathrm{IN}}$ values have Fe$_{\mathrm{OUT}}$ values which are comparable to the rest of the sample at 0.30 and 0.29 respectively, very close to the median value of 0.23. This again highlights how metallicity is thought to be relatively uniform outside of the cluster core, even for clusters which are noticeably disturbed in the core.

\subsection{Median silicon \& oxygen profiles}

Beyond the iron and total metallicity profiles, it is also instructive to look at the profiles of other metals which are produced through other astrophysical processes. Fe is predominantly formed in supernovae Type Ia (SNIa, explosions of binary white dwarfs), whereas elements such as silicon and oxygen are thought to be mainly produced in core-collapse supernovae (SNII). Assuming that enrichment via SNIa and SNII occurs on different timescales, it can be useful to compare profiles from the different enrichment channels to probe the relative abundance of different star burst events and gain insight into where they are likely to occur within a cluster.

Fig.~\ref{fig:Si_O_comparison} shows the the median mass-weighted Si (green) and O (orange) profiles (left panel). The $1~\sigma$ scatter in the clusters is shown by the coloured bands. We also show observational data from \cite{Mernier2017}, using crosses for Si and circles for O, although some of the data points have such low metallicity we do not show them on this plot. In the right panel we give the median relative abundance profiles, Si/Fe and O/Fe, and the scatter using the same colour scheme. These profiles were obtained by first calculating the relative abundance profiles for the individual clusters and then taking the median. The observational data for Si/Fe is taken from observations of AWM7 \citep{Sato2008}, Centaurus \citep{Sakuma2011} and Coma \citep{Matsushita2013}. The observations of AWM7 and Centaurus also provided measurements of O/Fe. 

\begin{figure*}
	\includegraphics[width=\textwidth, trim=3cm 2.5cm 2.5cm 2.5cm, keepaspectratio=True]{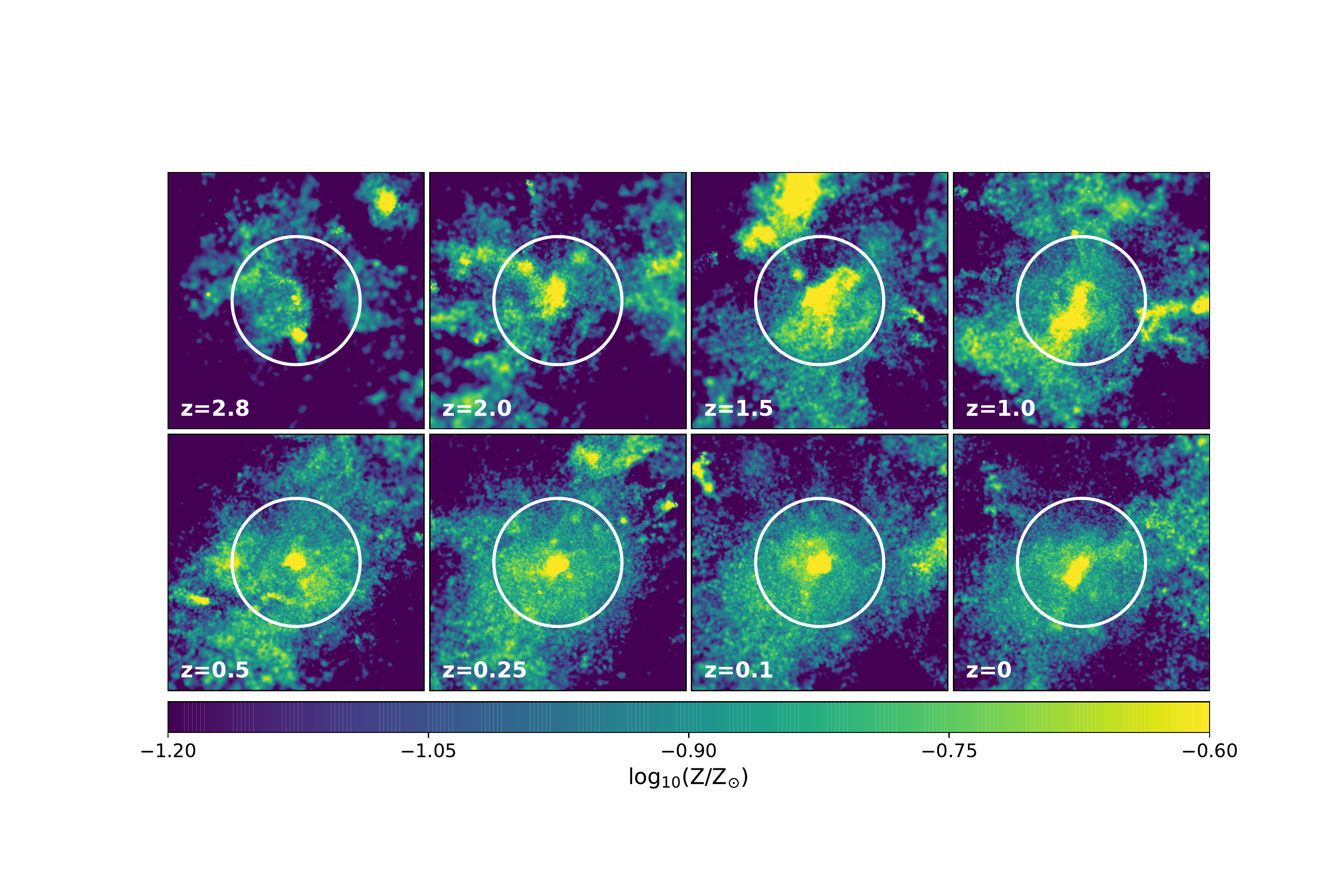}
	\caption{The metallicity maps of CE-05 from $z=2.8$ to $z=0$. All of the maps have an extent of $4~r_{500}$ in all directions and are projected down the $z$-axis, while the white circle shows the position of $r_{500}$. The central metallicity peak has formed by $z=2$, with very little evolution in the cluster from at least $z=0.5$ to present.}
	\label{fig:map_z}
\end{figure*}

\begin{figure*}
	\includegraphics[width=\textwidth,keepaspectratio=True,trim=0cm 1cm 0cm 0cm]{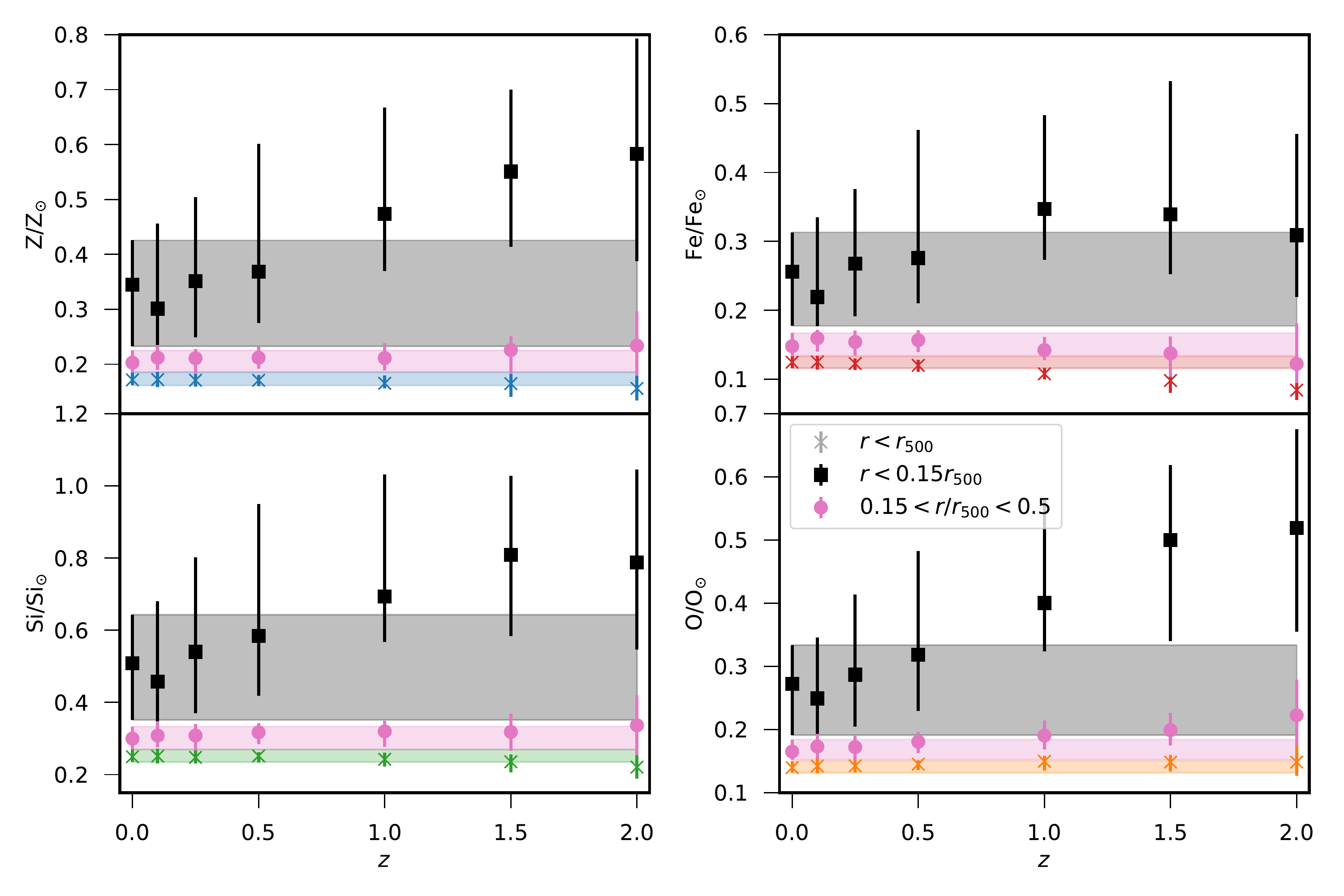}
	\caption{The redshift evolution of mass-weighted metals within three different radial regions: $r < 0.15~r_{500}$ (black squares), $0.15 < r/r_{500} < 0.5$ (pink circles) and $r < r_{500}$ (coloured crosses). The top row shows total metallicity and silicon, while the bottom row shows iron and oxygen. The scatter in the data at each redshift is shown by the error bars on individual points. The shaded regions show the 1~$\sigma$ scatter at $z=0$ for each of the radial ranges in the same colour as the individual points.}
	\label{fig:metals_z}
\end{figure*}

From the left panel, we can see that our simulations produce levels of silicon and oxygen that are very consistent with observations. The final data points of both the Si and O profiles obtained by \cite{Mernier2017} have considerably lower Si/Si$_{\odot}$ and O/O$_{\odot}$ values than at smaller radii, such that we do not show the final data point for the O profile because it is outside of the plotted range, O/O$_{\odot} = 6\times 10^{-3}$. This sharp decline in the observed profiles highlights the current difficulty in measuring metallicity at large radii, something which upcoming X-ray telescopes are expected to improve on. It also suggests that the metallicity profiles obtained from different enrichment channels could show differing trends at large radii. Fe, which is primarily produced by SNIa, is currently expected to plateau for $r > 0.3~r_{500}$ (see \citealt{Mernier2018} and references therein), while metals produced mainly though SNII seem to show a gradual radial decline.

In the right panel, the relative abundance profiles for the C-EAGLE sample lie above the observational data. As the individual silicon and oxygen profiles are in such good agreement with observations this is likely driven by the Fe abundance being too low (see Fig.~\ref{fig:Z_Fe_comparison}), leading to an increase in normalisation of the relative abundance profiles. Indeed, if we rescale the Fe profile by a factor of 1.5, needed to bring the Fe profiles into agreement with the observations, the normalisation of the Si/Fe and O/Fe profiles decrease such that they are in very good agreement with the observations (not shown). The shape of the relative abundances profiles agrees relatively well with the observations which seem to suggest a flat profile out to $r_{500}$, indicating that the ratio of metals from different SNe should not change with radius. In order to increase the Fe abundance in the C-EAGLE simulations whilst maintaining good agreement between the relative number of SNIa and SNII, the Fe yield would likely need to be increased without changing the supernova rates. Both the relative abundance profiles obtained from the simulations are very flat out to $r_{500}$, with only a slight increase between $1-7~r_{500}$. This suggests that there is a marginal increase in metals from SNII relative to those from SNIa for $r > r_{500}$ compared to within $r_{500}$. There is considerably more scatter in the core of the relative abundance profiles than at any other radii. This is driven by the increased scatter in the median of the Fe, Si and O profiles at small radii (left panel and Fig.~\ref{fig:Z_Fe_comparison}). 

\section{Evolution of the ICM}\label{Sec:Redshift}

In this section, we will look at the redshift evolution of metals in the C-EAGLE sample. Specifically, we will compare the different metallicity production channels (AGB stars, SNIa and SNII) to see how their contribution to the overall metal distribution changes with time, and look at how the cluster cores form.

In Section~\ref{Sec:PresentDay} we presented the present day metallicity distribution in the C-EAGLE clusters. Outside of the cluster cores, $r > 0.15$, we have shown that the metal abundance in different radial apertures is remarkably constant across the entire sample and does not sensitively depend on any observable cluster property that was tested. The metallicity profiles show that the metal abundance decreases at larger radii, with very little scatter outside of the cluster cores. In order to explain this apparent homogeneity of metallicity, both within the outskirts of individual clusters and across the cluster population, the `early enrichment' picture of metal abundance has been developed in which the majority of metals are produced at high redshift, $z > 2$, in the cores of cluster progenitors (e.g. \citealt{Mantz2017, Mernier2018}).

\begin{figure*}
	\includegraphics[width=\textwidth,keepaspectratio=True,trim=0cm 1cm 0cm 0cm]{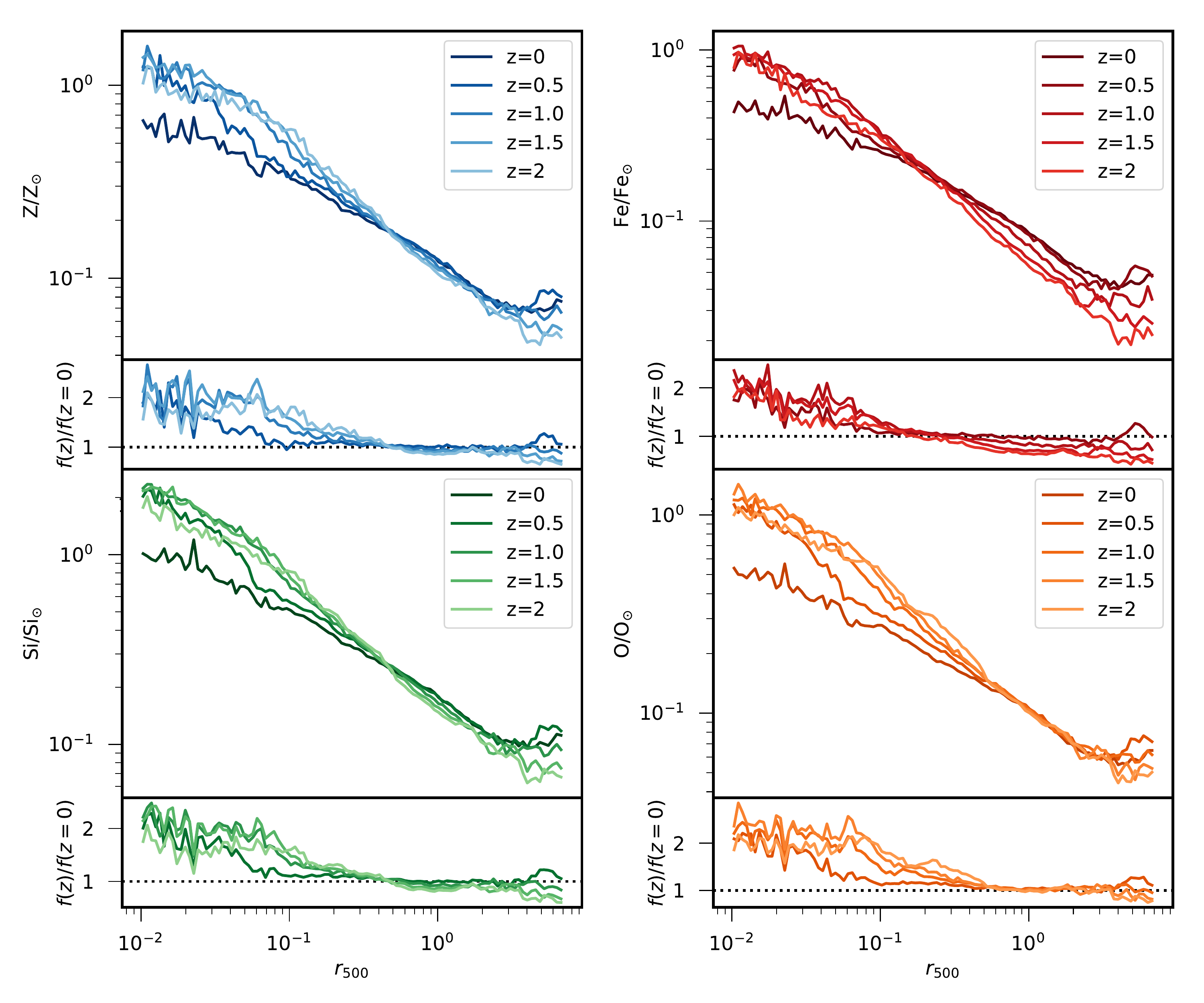}
	\caption{The redshift evolution of the median, mass-weighted metallicity profiles for Z, Fe, Si and O, scaled by their respective solar abundances. In each panel the lighter lines correspond to higher redshifts. In each subplot we show each profile at $z > 0$ divided by the $z=0$ profile to highlight any redshift evolution in the metal abundance. For all of the different metals there is significant evolution in the core of the clusters, $r < 0.4~r_{500}$, between redshifts 0.5 and 0. Iron is the only profile which shows any evolution with time at large radii.}
	\label{fig:metal_profile_z}
\end{figure*}

In Fig.~\ref{fig:map_z}, we present metallicity maps for a single cluster, CE-05, from $z=2.8$ to $z=0$. It can be seen that by $z=2$, the cluster progenitor already shows evidence for a high metallicity peak in the core which is maintained until the present day. Between $z=2$ and $z=1$, the relative size and shape of the metallicity peak changes due to the accretion of high metallicity structures from outside of $r_{500}$ (seen in the top left of $z=1.5$ panel). Since $z=0.5$, there is essentially no evolution in either the core of the cluster or in the outskirts. At $z=0$, CE-05 is an ongoing merger and is classified as unrelaxed.

\subsection{Evolution of metallicity fractions}

Fig.~\ref{fig:metals_z} shows the redshift evolution of mass-weighted metallicity in three different apertures; $r < 0.15~r_{500}$ (black squares), $0.15 < r/r_{500} < 0.5$ (pink circles), and $r < r_{500}$ (coloured crosses). The median value in each radial region is plotted at different redshifts, with the errorbars showing the $1~\sigma$ scatter in the sample at each epoch. The shaded regions show the scatter in the C-EAGLE clusters at $z=0$ to highlight any redshift evolution in the data. We present data for the total metallicity (blue), Fe (red), Si (green) and O (orange) abundances.

\begin{figure*}
	\includegraphics[width=\textwidth,keepaspectratio=True,trim=0cm 1cm 0cm 0cm]{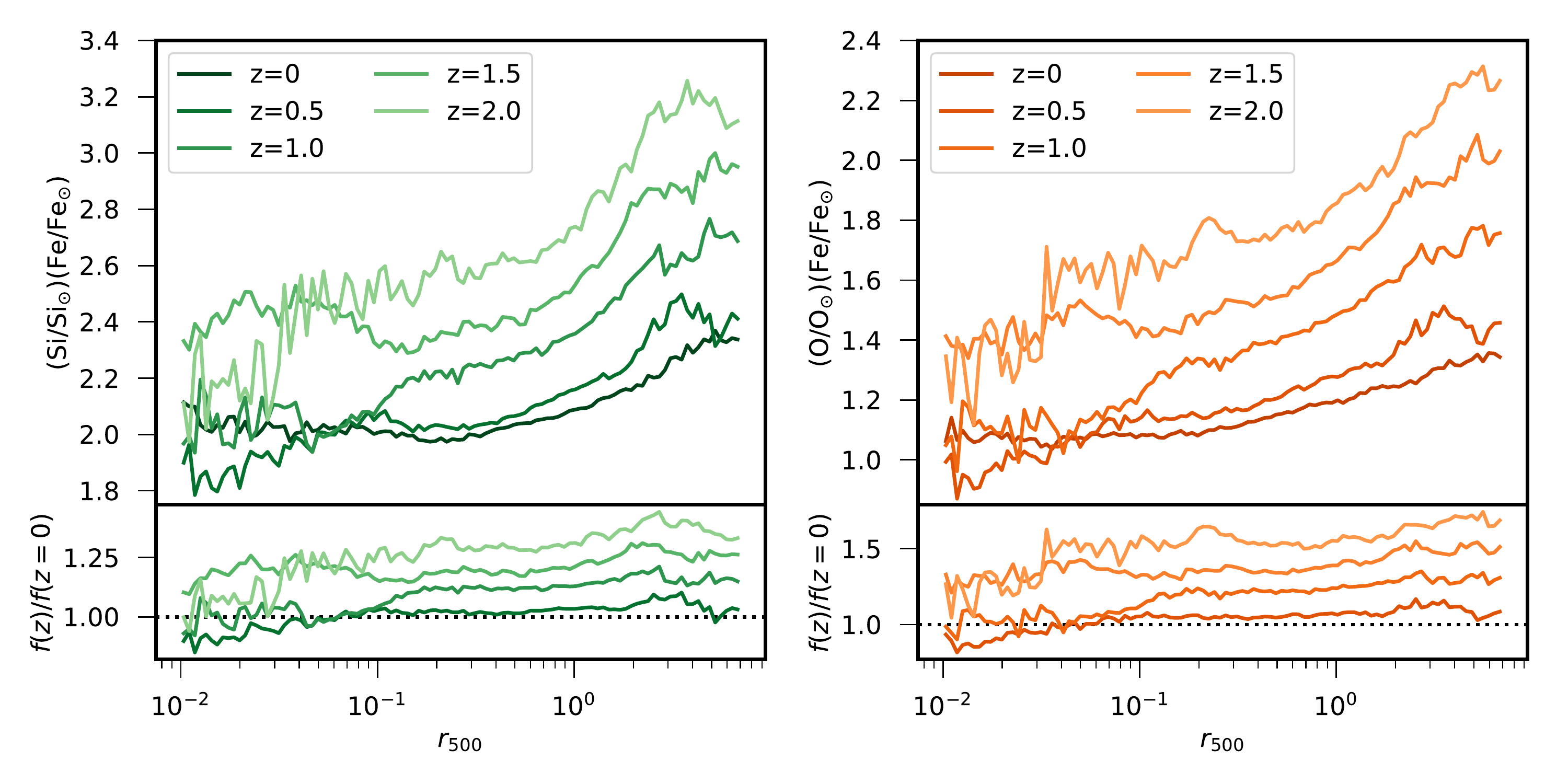}
	\caption{Redshift evolution of the abundance ratios for silicon (left panels) and oxygen (right panels) with respect to Fe. Lighter lines correspond to higher redshift, with the darkest line giving the $z=0$ profile (as seen in Fig.~\ref{fig:Si_O_comparison}). The bottom panels for both metals show the abundance ratios compared to the $z=0$ profile. For both metals, any evolution in the core happens between $z=1.5$ and $z=1$, with the cluster outskirts ($r > 0.1~r_{500}$) not becoming self-similar until $z=0.5$. Below $z=0.5$, there is essentially no evolution in the abundance ratios.}
	\label{fig:abundance_profile_z}
\end{figure*}

For all of the metals presented, the scatter in the abundance within $r_{500}$ (coloured crosses) is very small at all redshift. Typically, the scatter is less than 10 per cent of the median for $z < 1$ and increases only slightly to around 20 per cent for $z=1.5$. In contrast, the minimum scatter in the smallest central aperture ($r < 0.15~r_{500}$) is at least 25 per cent of the median value. This decrease in scatter when moving to larger radii is also seen by \cite{Biffi2017} when looking at the Fe abundances in their simulated sample. Our median Fe/Fe$_{\odot}$ values for the C-EAGLE sample are low compared to those found by \cite{Biffi2017} at all redshift for all of the different apertures, again highlighting how our simulations are Fe deficient (see Fig.~\ref{fig:Z_Fe_comparison}).

Focussing on the cluster cores, $r < 0.15~r_{500}$, we find that there is considerable redshift evolution in the metal abundances, which decrease towards low redshift. This is in direct contention with previous studies which have found no evolution of core metallicity \citep{Biffi2017, Vogelsberger2018}, and observational studies which typically find an anti-correlation between central metal abundance and redshift \citep{Ettori2015, McDonald2016, Mantz2017, Liu2018}. When selecting only the NCC clusters in their respective samples, \cite{Ettori2015} and \cite{McDonald2016} find no evolution of the metal abundance, suggesting that any measured redshift dependence of metallicity in the core is driven by CC clusters. Under this assumption, as the C-EAGLE sample is known to include no CC clusters (or 4 depending on the definition of CC used), it would be expected that there would be no redshift evolution in the C-EAGLE clusters. However, when separating their cluster sample with respect to peakiness, \cite{Mantz2017} find that those clusters with low peakiness evolve most with redshift. As high cluster peakiness is assumed to be an indicator of cool-coreness, this is at odds with the findings of other observations. The decrease in metallicity in the core of the C-EAGLE clusters at low redshift is discussed in more detail in Section~\ref{Sec:Core}.

At larger radii, $0.15 < r/r_{500} < 0.5$, there is a mild anti-correlation between Fe abundance and $z$, although the data are consistent with no evolution within the scatter. This is similar to the result of \cite{Ettori2015} who also found evidence for decreasing metallicity at high redshift in a similar aperture when considering the whole cluster sample, and no evolution when taking the NCC subset. \cite{McDonald2016} found both the CC and NCC samples were consistent with no redshift evolution at large radius, while \cite{Mantz2017} again found a potentially contradictory result that only the low peakiness clusters evolved with redshift. To determine if the different observations are in contention, a study is needed to determine how peakiness correlates with cool-coreness, specifically in the \cite{Mantz2017} cluster sample, although due to the definition of cool-core clusters it is assumed that they would correlate very strongly.

Within $r_{500}$, the abundances of Z, Si and O are all consistent with no redshift evolution. There is again a mild anti-correlation between Fe and $z$, which is significant at $z=1.5$ compared to the scatter at $z=0$. Fe is typically produced through SNIa explosions, whereas the bulk of Si and O are released through SNII. The difference in the metal abundances at large radii is likely a result of the different metal production channels, which we discuss in more detail in Sections~\ref{Sec:Core} and \ref{Sec:mass}.

\subsection{Evolution of metallicity profiles}

Fig.~\ref{fig:metal_profile_z} shows the redshift evolution of the median, mass-weighted metal abundance profiles. The four panels show the total metallicity (blue), Fe (red), Si (green) and O (orange) profiles, with the darker coloured lines corresponding to low $z$. The abundance profiles are shown between $z=0$ and $z=2$ for all of the metals. In the subplots we also give the ratio of the profiles compared to the $z=0$ profile.

Regardless of the metal there is significant redshift evolution in the core, $r < 0.4~r_{500}$, of all the profiles. Typically the metallicity in the very centre of the cluster is at least a factor of 2 greater than the present day value at $z=1.5$. In the C-EAGLE clusters, the strongest evolution seems to occur between $z=0$ and $z=0.5$, with very little difference in the metallicity profiles for $z > 0.5$ until $z =2$. This increase in metallicity in the core is likely to be due to either increased AGN activity at high redshift leading to cluster cores with fractionally less gas than at $z=0$, allowing for a higher metallicity fraction, or there could be more low metallicity gas in the core at low redshift. We discuss these possibilities further in Section~\ref{Sec:Core}.

At larger radii, $r > 0.4~r_{500}$, there is essentially no redshift evolution in the total metallicity, Si and O profiles until $3~r_{500}$. However, the Fe abundance profile decreases more sharply at large radii for $z > 0.5$. This behaviour is also seen in the IllustrisTNG clusters (see fig.~11 of \citealt{Vogelsberger2018}), but no comment is made by the authors. As Fe is mainly produced via SNIa while Si and O are used as proxies for SNII, it suggests that the difference seen in the profiles is due to the differing evolution of the metal production channels. As SNIa are explosions of white dwarf binary systems, it takes longer for them to form compared to SNII which are the explosions of short-lived, massive stars. SNII are rare at low redshift as a result of the star formation rate, so the majority of massive stars are at large radii will already have exploded compared to an excess of white dwarf binaries which have yet to explode, leading to an evolution in the outskirts of Fe abundance profile. As the total metallicity profile does not show any evolution, it shows that Fe is not the main contribution to Z. For all of the metals, the profiles again seem to plateau at $3~r_{500}$ for all redshift.

Fig.~\ref{fig:abundance_profile_z} shows the redshift evolution of the relative abundance profiles of Si/Fe (left, green) and O/Fe (right, orange). Any evolution in the profiles suggests differences in the abundance of the different metal production channels, specifically SNIa (Fe) and SNII (Si, O). Lighter lines correspond to higher redshift, while the darkest line is $z=0.$ The bottom panels in each subplot show the ratio of each abundance profile compared to $z=0$, highlighting any redshift evolution of the metal abundance. 

Within the core, $r < 0.1~r_{500}$, there is very little evolution in either metal abundance profile until $z=1$ when the value of Si/Fe (O/Fe) increases by 0.2~Si$_{\odot}$/Fe$_{\odot}$ (0.4~O$_{\odot}$/Fe$_{\odot}$) to $z=1.5$. This suggests a relative increase in Si and O compared to Fe, so either a decrease in the number of SNIa or an increase in SNII. We explicitly compare the importance of different metal production channels in Fig.~\ref{fig:mass_fractions_profile_z} below. 

Outside of the cluster core ($r > 0.1~r_{500}$) there is evidence for evolution of the profile between $z=1.5$ and $z=0.5$. During this time the relative abundance decreases similar to at small radii, again suggesting that there is an increase in SNIa relative to SNII which happens at lower redshift than in the core. This trend of decreasing relative abundance at low $z$ is also found by \cite{Biffi2017} when looking at 4 clusters between $z=2$ and $z=0$. They find that the strongest evolution happens between $z=2$ and $z=1$, highlighting how there is very little change in the relative abundance profiles at low redshift.

\begin{figure}
	\includegraphics[width=\columnwidth,keepaspectratio=True,trim=0cm 1cm 0cm 0cm]{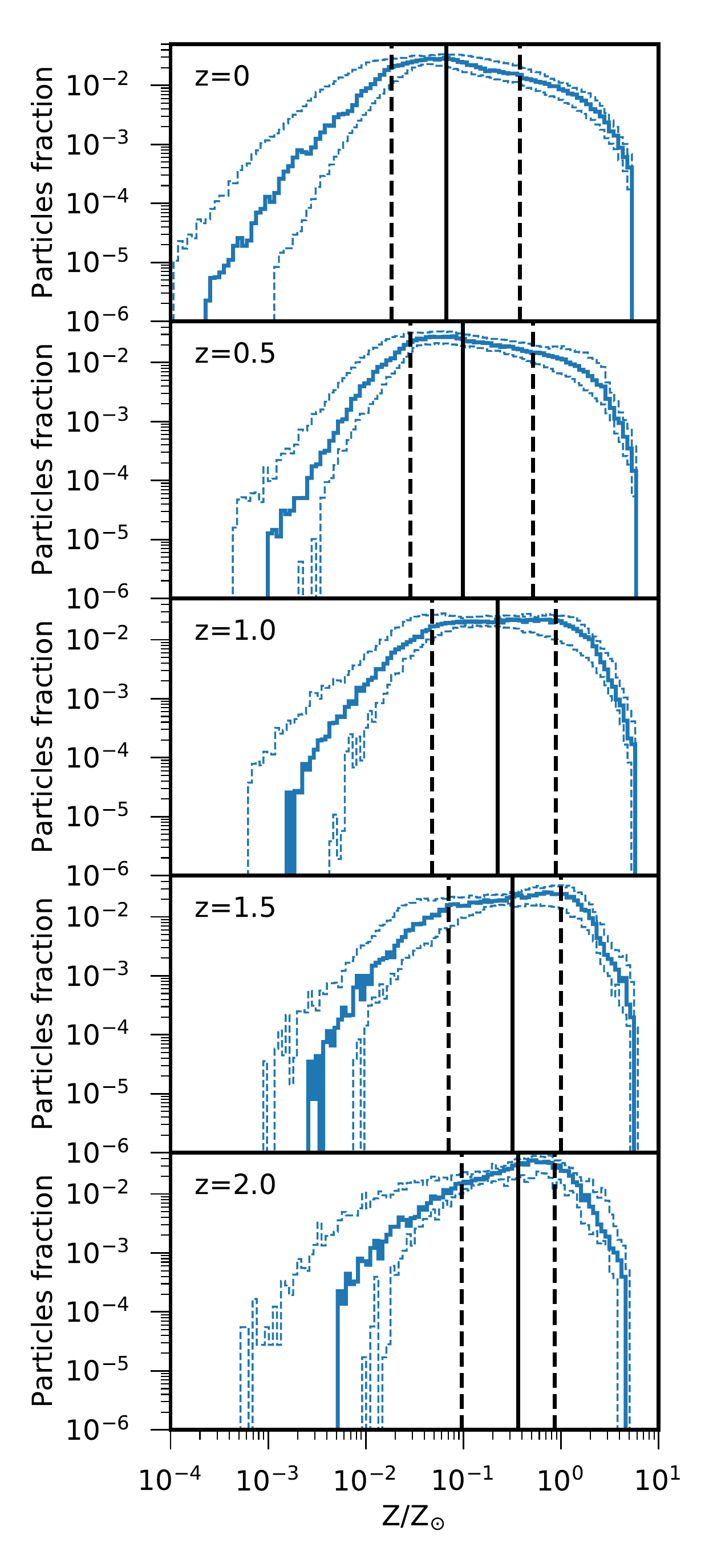}
	\caption{Histograms of the mass-weighted metallicity distribution in the C-EAGLE sample at different redshifts. The thick blue line shows the median distribution of individual gas particles across the whole sample, which the dashed blue lines show the $1~\sigma$ scatter in each metallicity bin. The median value of metallicity at each redshift is given by the vertical black line, with the 16th and 84th percentiles given by the dashed black lines. We normalise all of the individual metallicity histograms before taking the median to remove the impact of comparing clusters with different masses and different numbers of gas particles. We use all particles in the radial range $r < 0.1~r_{500}$ in all clusters and at each redshift. The median metallicity in the cluster cores can be seen to decrease at low redshift.}
	\label{fig:metals_dist_z}
\end{figure}

\subsection{Origin of cluster cores}\label{Sec:Core}

\begin{figure*}
	\includegraphics[width=\textwidth,keepaspectratio=True,trim=0cm 1cm 0cm 0cm]{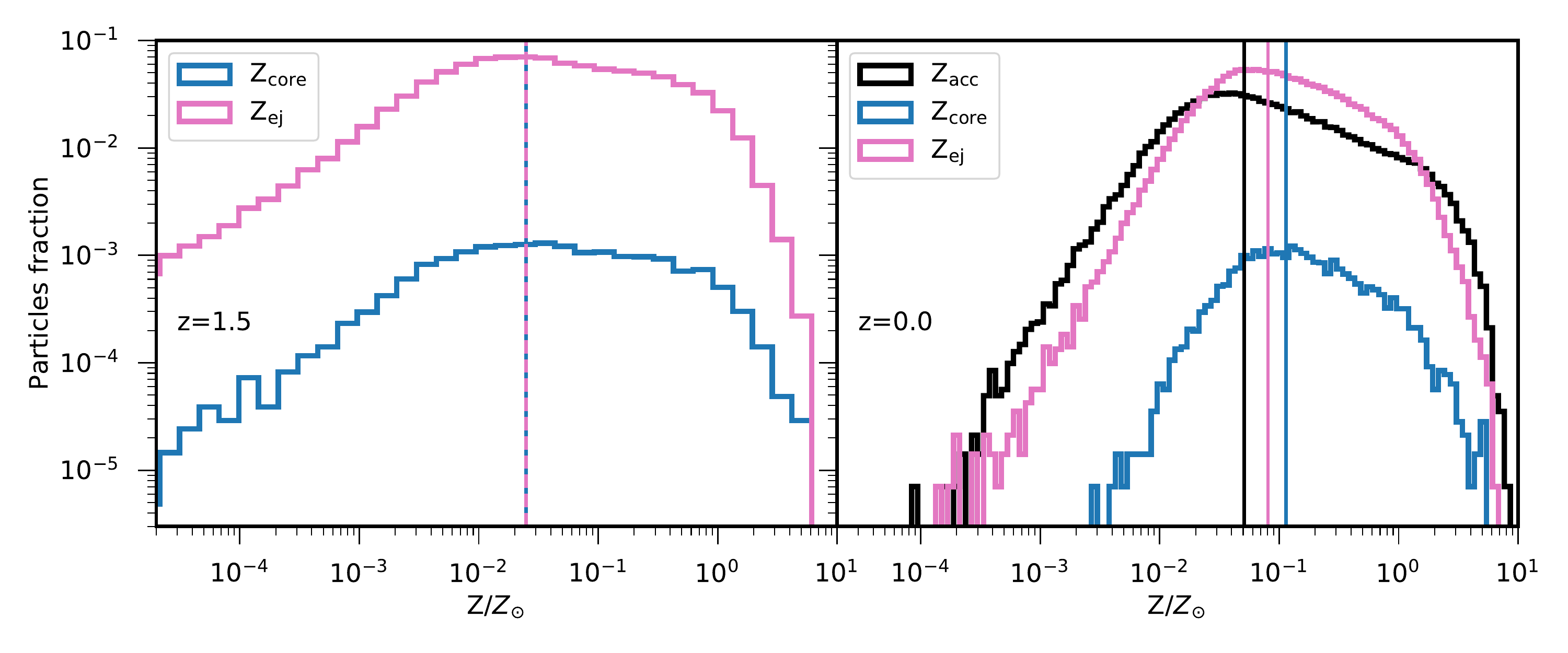}
	\caption{Histograms comparing the metallicity in the core, $r < 100$~kpc, at $z=1.5$ and $z=0$ for a single halo, CE-08. At $z=1.5$ we isolate all the particles in the region of interest and then track these particles forward to $z=0$ and see whether they remain inside the core, $Z_{\mathrm{core}}$, or if they are ejected from the core and are found at larger $r$ at $z=0$, $Z_{\mathrm{ej}}$. In the left panel we present the metallicity of the two samples at $z=1.5$, while in the right panel we show how the metallicity has evolved by $z=0$. At $z=0$ we present a third histogram which gives the metallicity of the gas which has been accreted into the core between the two epochs, $Z_{\mathrm{acc}}$. The vertical lines show the median of the distribution of the same colour.}
	\label{fig:CE_08_core}
\end{figure*}

In this subsection we look in more detail at the metallicity evolution in cluster cores. For the C-EAGLE clusters we have shown a positive correlation between the metal abundance in cluster cores and redshift (Fig.~\ref{fig:metals_z}) in contrast with other analyses which have found either no evolution \citep{Biffi2017, Vogelsberger2018} or a mild anti-correlation in observed data \citep{Ettori2015, McDonald2016, Mantz2017, Liu2018}. In Fig.~\ref{fig:metals_dist_z} we present histograms of the total metallicity distribution within 0.1~$r_{500}$ at different redshifts. The thick, solid blue line shows the median metallicity distribution, while the thin, dashed blue lines are the histograms obtained by taking the 16th and 84th percentiles within each metallicity bin. At each redshift the median of the different distributions is shown by the black vertical lines, with the dashed lines again corresponding to the percentile histograms, showing the approximately 1~$\sigma$ scatter in the data. Individual cluster histograms are normalised by the number of particles in the core before the median and percentiles are taken.

As expected, there is a decrease in median metallicity between $z=2$ and $z=0$ in the chosen aperture, $r < 0.1~r_{500}$. This decrease appears to be driven by the formation of a low metallicity tail in the core at low redshifts. This tail is likely caused by the accretion of low metallicity gas into the cluster core, but the overall metallicity distribution can also be effected by a high star formation rate leading to the removal of high metallicity gas as it has the shortest cooling time, or through the ejection of high metallicity gas as a result of AGN feedback, or through the accretion of low metallicity gas into the core. We compare these options in Fig.~\ref{fig:CE_08_core} for a single cluster, CE-08. For this analysis, we have isolated all of the particles at $z=1.5$ within 100~kpc of the cluster centre and tracked each particle to its final position at $z=0$. From this, we were able to separate the particles at $z=1.5$ into those that remain in the core at $z=0$, Z$_{\mathrm{core}}$, and those which are ejected between the two epochs, Z$_{\mathrm{ej}}$. Similarly, at $z=0$, we can introduce a third sample of particles which were outside of the core at $z=1.5$, but which have been accreted into the core by $z=0$, Z$_{\mathrm{acc}}$. We found that the fraction of gas particles forming stars was negligible so we do not include this in our analysis. We chose to use particles within 100~kpc for this analysis instead of $0.1~r_{500}$ due to needing a large enough aperture at $z=1.5$ to enable enough particles to be in the cluster core but also not resulting in too many particles at $z=0$ due to the computational intensity of tracking the individual particles.

In Fig.~\ref{fig:CE_08_core}, the left panel shows the metallicity histograms of Z$_{\mathrm{core}}$ (blue line) and Z$_{\mathrm{ej}}$ (pink line) at $z=1.5$. Summing the two histograms together would give the total metallicity distribution in the core of CE-08 at $z=1.5$. The right panel shows the histograms of Z$_{\mathrm{core}}$ (blue line), Z$_{\mathrm{ej}}$ (pink line) and Z$_{\mathrm{acc}}$ (black line) at $z=0$. Summing the Z$_{\mathrm{core}}$ and Z$_{\mathrm{acc}}$ lines together would give the total metallicity distribution of the core at $z=0$. The vertical lines show the median of the distribution of the corresponding colour. 

At $z=1.5$, the median of both Z$_{\mathrm{core}}$ and Z$_{\mathrm{ej}}$ are the same, but by $z=0$ the median metal abundance of Z$_{\mathrm{core}}$ has increased beyond that of Z$_{\mathrm{ej}}$. Comparing the three samples at $z=0$, the particles that remain in the core have the highest metallicity, as expected from Fig.~\ref{fig:metal_profile_z}. However, Z$_{\mathrm{ej}}$ has higher median metal abundance than Z$_{\mathrm{acc}}$ which ties into the early enrichment model as it suggests that gas is preferentially enriched in the core of (proto-)clusters at high redshift. Although the Z$_{\mathrm{acc}}$ gas sample is in the cluster core at $z=0$, it has lower metallicity than the Z$_{\mathrm{ej}}$ sample because there is less enrichment of gas by stars at low redshift. It can also be seen that the total metallicity increases in the core as a whole between $z=1.5$ and $z=0$. This is in contrast with what is expected from Figs.~\ref{fig:metal_profile_z} and \ref{fig:metals_dist_z}, and is caused by the use of a radial range fixed in physical coordinates, $r < 100$~kpc, rather than using co-moving coordinates. When comparing apertures of $r < 0.1~r_{500}$, the metallicity decreases between $z=1.5$ and $z=0$ as expected due to the increase in low metallicity gas.

Between $z=1.5$ and $z=0$, only around 2 per cent of gas remains in the cluster core. Although we have shown that the accretion of low metallicity gas causes the core metal abundance to decrease, it is also clear that AGN feedback is very important in producing this result. Between the two epochs the majority of the gas in the cluster core has been removed as a result of bulk motions in the cluster core which are driven by AGN feedback. In order to reproduce the observational result of no evolution in cluster cores, it is likely that the strength of AGN feedback in C-EAGLE would need to be reduced. From Fig.~\ref{fig:CE_08_core}, the ejected gas has higher metallicity than the accreted gas, so if a percentage of Z$_{\mathrm{ej}}$ remained in the core at $z=0$, the metallicity of the core would increase. The importance of mixing the gas by large scale motion was also found by \cite{Lovisari2019} when comparing relaxed and unrelaxed clusters, building on the work by e.g. \cite{Biffi2017} who find that AGN feedback is crucial in reproducing observed metallicity trends. Although we only present this analysis of the core for one cluster, we have looked at the evolution of other clusters and found the same trends. 

\subsection{Metallicity mass fractions}\label{Sec:mass}

\begin{figure}
	\includegraphics[width=\columnwidth,keepaspectratio=True,trim=0cm 1cm 0cm 0cm]{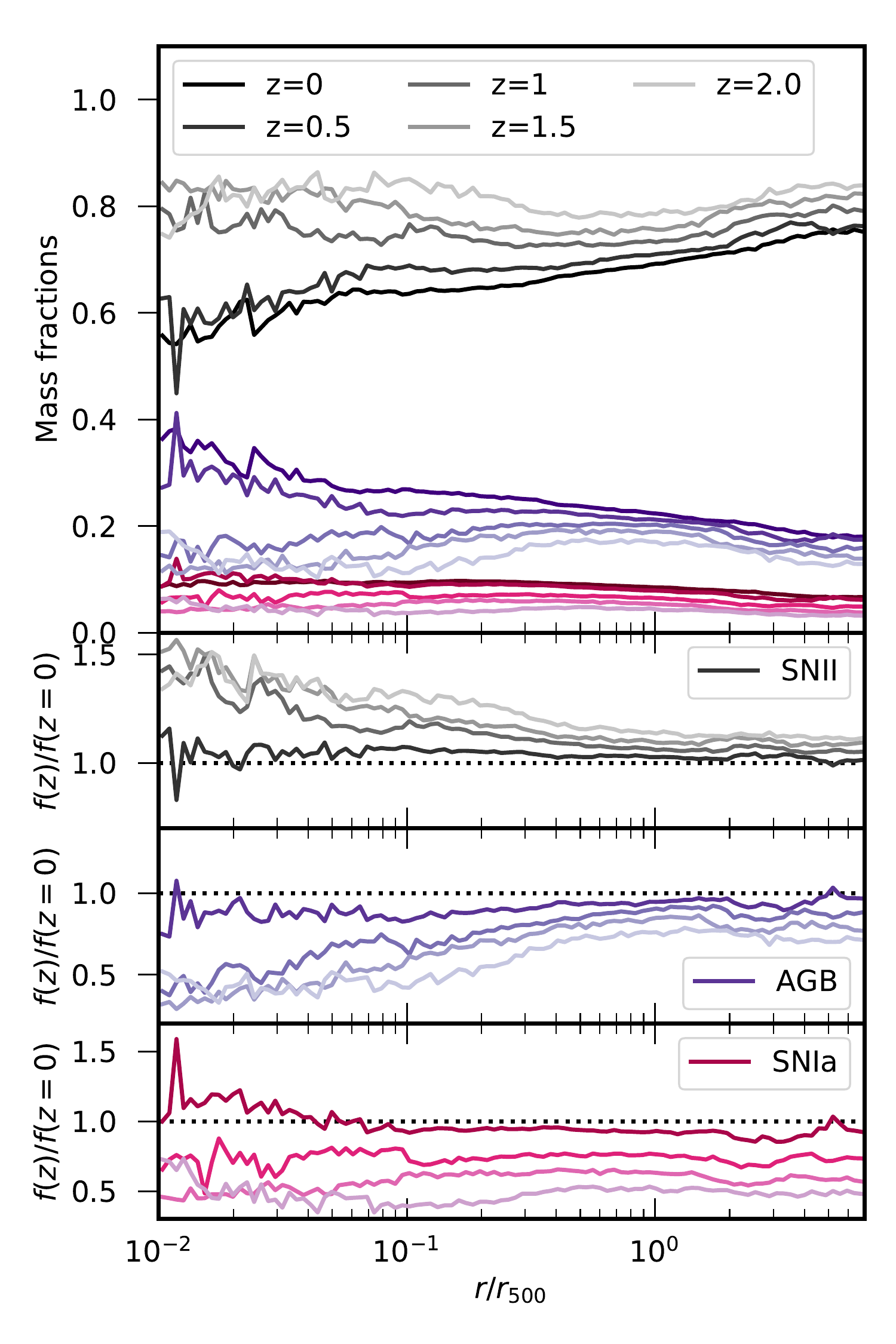}
	\caption{The fraction of mass produced by SNIa (pink), SNII (grey) and AGB stars (purple) as a function of radius from the cluster centre. Each line is the median mass fraction profile across the 30 C-EAGLE clusters at a given redshift. The darkest lines correspond to $z=0$, with lines getting lighter moving towards higher $z$ values. The three lower panels show the mass fractions of $z > 0$ compared to the $z=0$ profile for SNII, AGB stars and SNIa respectively. The strongest evolution in mass fractions can be seen to occur between $z=1.5$ and $z=0.5$ for all of the metal production channels.}
	\label{fig:mass_fractions_profile_z}
\end{figure}

In Fig.~\ref{fig:mass_fractions_profile_z} we present the mass fractions of total metallicity produced through SNIa (grey), SNII (pink) and AGB stars (purple). Lighter lines correspond to higher redshifts, with the lightest lines giving the radial distribution of mass from each production channel at $z=2$ while the darkest lines gives the profiles at $z=0$. The subplots show the ratio of the mass fraction profiles to the profile at $z=0$ to highlight any redshift evolution in the different channels. At each redshift, the sum of the combined SNIa, SNII and AGB mass fractions at a given radius is equal to unity.

At all redshifts, the contribution of metals from SNII is the largest while SNIa contribute the least, as expected. The contribution from SNII is seen to increase at high-$z$, while the mass fractions from SNIa and AGB stars increase at low-$z$. These trends are also found by \cite{Biffi2017} and \cite{Vogelsberger2018} using different simulated samples, and is driven by the difference in lifetimes between the massive, short-lived stars which result in SNII, and low-mass long-lived stars which contribute to SNIa and AGB star populations. However, \cite{Biffi2017} find much higher levels of enrichment from SNIa in their clusters with the mass fractions increasing from around a quarter at $z=2$ to a third at $z=0$, compared to a maximum contribution of around 10 per cent found in the C-EAGLE clusters and the IllustrisTNG simulations \citep{Vogelsberger2018}. This difference can likely be attributed to the assumption of different supernova rates and elemental yields in the different simulations, and explains the good agreement found by \cite{Biffi2017} to the observed Fe abundances. Observationally, \cite{Ezer2017} find that the number of SNIa is around 12-16 per cent of the total number of supernova explosions out to the virial radius.

\begin{figure}
	\includegraphics[width=\columnwidth,keepaspectratio=True,trim=0cm 1cm 0cm 0cm]{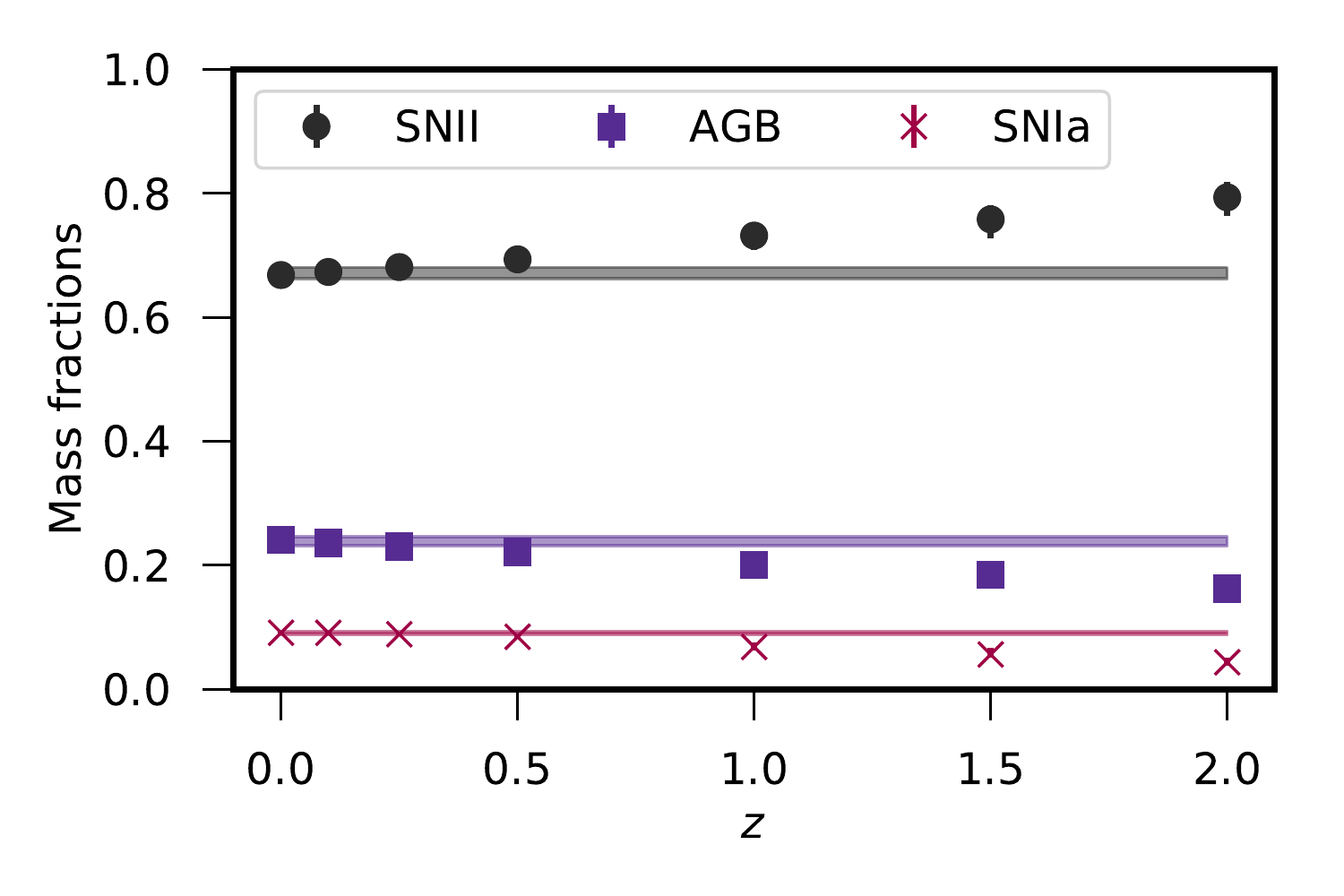}
	\caption{The median mass fractions within $r_{500}$ for the three different metal production chanels, SNIa (pink), SNII (grey) and AGB stars (purple), at different redshift. The shaded regions show the scatter in the mass fractions at $z=0$. Points are plotted with errorbars which give the 1~$\sigma$ scatter in the data at each redshift, but the scatter is typically less than 5 per cent so the errorbars are not visible.}
	\label{fig:mass_fractions_z}
\end{figure}

Looking at the radial dependence of the different metal production channels, the SNIa contribution is relatively constant at each redshift for the C-EAGLE clusters, but is found to increase at small radii by \cite{Biffi2017} and \cite{Vogelsberger2018}. Similar to \cite{Vogelsberger2018}, we find that the contributions from SNII and AGB stars are anti-correlated at all redshift, but the mass fraction profiles for the C-EAGLE clusters flatten at high redshift which is not seen in the IllustrisTNG simulation. For $z < 1.0$, the SNII contribution increases with radius, while the mass fraction from AGB stars decreases at large radius. Conversely, \cite{Biffi2017} find that the SNII mass fractions increase at small radii. This discrepancy is again likely due to the different assumptions and models used in the simulations.

For $z < 0.5$, the shape of the relative abundance profiles (Fig.~\ref{fig:abundance_profile_z}) are well recovered by taking a simple ratio of the mass fractions profiles from SNII and SNIa, assuming that the majority of Si and O are produced by SNII, and Fe by SNIa. However, at $z > 1.0$ this simple approach leads to an increase in relative abundance in the core, $r<0.1~r_{500}$, which is not seen in the C-EAGLE clusters. This suggests that at high redshift the other abundance channels (SNII and AGB stars) produce relatively more Fe at small radii than at low redshift in order to bring the Si/Fe and O/Fe abundance profiles down in the core. This trend of decreasing Fe mass fraction from SNIa at high redshift is also seen in \cite{Vogelsberger2018}, and is a result of fewer stars having formed at high redshift.

The strongest evolution in the mass fractions happens between $z=0.5$ and $z=1$. To test whether this evolution is significant we present the median mass fractions within $r_{500}$ for the three stellar populations at different redshift in Fig.~\ref{fig:mass_fractions_z}. SNIa are given by pink crosses, SNII by black circles and AGB stars by purple squares. All the datapoints are plotted with errorbars which given the 1~$\sigma$ scatter in the C-EAGLE clusters, but this is typically less than 5 per cent of the value so the errorbars are not visible. To highlight any redshift evolution in the mass fractions, we show the dispersion in the data at $z=0$ as the shaded region for each metal production channel, which again highlights how small the scatter is. There is a mild evolution between mass fraction and redshift for all three stellar populations, with the fraction of total metals from SNII decreasing at low $z$ and increasing from SNIa and AGN stars. It is clear that for $z > 0.5$, the evolution in mass fractions is significant as the datapoints lie outside of the expected scatter. That there is very little evolution in the mass fractions for $z < 0.5$ suggests that the star formation rate at low $z$ is very small, with very few metals being produced as expected.

The strongest evolution in the mass fractions happens between $z=0.5$ and $z=1$. To test whether this evolution is significant or if there is a lot of scatter in the mass fractions, we present Fig.~\ref{fig:mass_fractions_z}. Here, the median mass fraction within $r_{500}$ is given for the three different stellar populations, SNIa (pink crosses) SNII (black circles) and AGB stars (purple squares) at each redshift. All the datapoints are plotted with errorbars which give the $1~\sigma$ scatter in the C-EAGLE clusters, but this is typically less that 5 per cent of the value so the errorbars are not visible. To highlight any redshift evolution in the mass fractions, we show the dispersion in the data at $z=0$ as the shaded region for each metallicity channel, again highlighting how small the scatter in the data is. There is a mild evolution between mass fraction and redshift for all three stellar populations, with the fraction of total metals from SNII decreasing at low $z$ and increasing from SNIa and AGB stars.

\section{Summary \& future work}\label{Sec:Conclusions}

In this paper we have the looked at the metallicity distribution and evolution in the C-EAGLE simulated cluster sample. We have removed CE-27 from our analysis due to its unphysical evolution as a result of a large AGN feedback event at very high redshift. Our main results are summarised as follows:
\begin{enumerate}
	\item There is a mild anti-correlation in the mass - metallicity relation, Fig.~\ref{fig:Fe_M500}, in agreement with observational \citep{Yates2017} and simulation-based studies \citep{Truong2019}. We find most evidence for this in the core of the clusters, $r < 0.15~r_{500}$, however, the scatter in this aperture is greater than at larger radii. 
	\item The C-EAGLE clusters are Fe deficient leading to an offset in the normalisation of the Fe profile compared to observations, Fig.~\ref{fig:Z_Fe_comparison}. However, the overall shape of the profile is in very good agreement with observations \citep{Leccardi2008, Mernier2017} out to $0.5~r_{500}$ where the C-EAGLE clusters continue to decline, whereas the observations suggest the profile should flatten.
	\item We observe a bimodal distribution in the Fe profiles, Fig.~\ref{fig:Fe_profile}, despite the lack of cool-core (CC) clusters in the C-EAGLE simulations \citep{Barnes2017b}. To explain this bimodality, we follow \cite{Lovisari2019} and separate the clusters according to their relaxation state, with relaxed clusters showing the increasing metallicity profiles towards the cluster centre while unrelaxed clusters tend to plateau. However, at high redshift the fraction of cool-core clusters is expected to increase (e.g. \citealt{Barnes2018}), while the relaxed cluster fraction should decrease. Using a different definition, the C-EAGLE clusters have 4 CC clusters, but they evolve in the same way as the rest of the sample.
	\item The Si and O profiles are in good agreement with the observational profiles of \cite{Mernier2017}, Fig.~\ref{fig:Si_O_comparison}. The Si/Fe and O/Fe relative abundance profiles are very flat out to $7~r_{500}$, in good agreement with the shape of recent observations but the normalisation of the profiles is too high due to the low Fe abundance in our simulations. As the profile shape is well recovered, this suggests that the relative abundance of SNIa and SNII is well matched to observations so the supernova rates should not be changed. Instead, the Fe yield in the C-EAGLE simulations would need to be increased in order to increase the Fe abundance.
	\item We find considerable redshift evolution in the core of the C-EAGLE clusters which has not been seen in other studies. At high redshift the C-EAGLE core values are higher than at $z=0$, whereas typically no evolution is measured \citep{Biffi2017, Vogelsberger2018}, or a mild decrease of metallicity towards higher redshift is found \citep{Ettori2015, McDonald2016, Mantz2017, Liu2018}. We attribute this decrease in metallicity at low redshift to the accretion of low metallicity gas onto the cluster core, coupled with the ejection of low metallicity gas, Fig.~\ref{fig:CE_08_core}. This mixing is a result of AGN feedback leading to large scale motions in the core, such that less than 5 per cent of the gas remains in cluster cores between $z=1.5$ and $z=0$.
	\item In the cluster outskirts, we find very little evolution of the Z, Si and O metal abundance profiles beyond $0.3~r_{500}$ since $z=2$, Fig.~\ref{fig:metal_profile_z}. This is in good agreement with the early enrichment model, which explains the observed homogeneity of cluster outskirts since $z \simeq 2-3$ as a result of the majority of gas enrichment happened in the core of proto-clusters at high redshift. However, we find significant redshift evolution in the outskirts of the Fe profile which is predominantly produced by SNIa, compared to Si and O which are typically produced by SNII. SNIa are long-lived low-mass stars which take longer to form and, therefore, to explode, so the Fe profile continues to evolve after the main epoch of star formation whereas SNII are massive, short-lived stars which explode relatively soon after formation.
	\item At all redshifts, the contribution to the overall metallicity from SNIa is less than 10 per cent at all radii (Fig.~\ref{fig:mass_fractions_profile_z}). The contributions from SNII and AGB stars are anti-correlated, with the mass from SNII decreasing at low redshift close to the cluster core while the mass contributed by AGB stars increases. This is again due to the difference in lifetimes between the stellar populations, with AGB stars continuing to produce metals after SNII due to their long lifetimes. The shape of the relative abundance profiles (Fig.~\ref{fig:abundance_profile_z}) is well recovered at low redshift simply by taking the ratio of SNII/SNIa. However, at high redshift the core is overestimated which suggests that the other abundance channels (AGB stars and SNII) produce relatively more Fe at small radii.
\end{enumerate}

Although the C-EAGLE cluster outskirts appear to be in good agreement with the early enrichment model, we have shown that there is evolution in the Fe profile since $z=2$ which is not seen observationally. We have also found a correlation between metal abundance in the core and increasing redshift, the opposite trend to what has been observed. The next generation of X-ray telescopes, such as \textit{Athena}, will be able to provide more measurements of Fe abundance at high redshifts which will improve constraints on the enrichment history and help to determine when most metals are released \citep{Pointecouteau2013,Cucchetti2018}.

In order to provide a better match to observations of Fe abundance whilst maintaining the observed profile shape and the good agreement with Si and O abundances, one option is to increase the Fe yield. However, the Fe yield is not well established observationally, so simply increasing the yields is not well motivated. Another option would be to increase the SNIa rates which would be expected to improve the Fe abundance but is likely to reduced the agreement with observations found for the shape of the Si/Fe and O/Fe relative abundance profiles. We have also established that reducing the strength of AGN feedback could be required to remove the redshift evolution of metal abundance found in the core of the C-EAGLE clusters. This could also help to increase the Fe abundance, but is also likely to change the global metal distribution in the clusters. Overall, reproducing the observed metallicity trends of clusters is a multi-faceted problem that relies on an in-depth understanding of both chemical enrichment and evolution, as well as galaxy formation, which should be improved by the next generation of X-ray missions.

\section*{Acknowledgements}
This work used the DiRAC@Durham facility managed by the Institute for Computational Cosmology on behalf of the STFC DiRAC HPC Facility (www.dirac.ac.uk). The equipment was funded by BEIS capital funding via STFC capital grants ST/K00042X/1, ST/P002293/1, ST/R002371/1 and ST/S002502/1, Durham University and STFC operations grant ST/R000832/1. DiRAC is part of the National e-Infrastructure. FAP is supported by an STFC studentship. YMB acknowledges funding from the EU Horizon 2020 research and innovation programme under Marie Sk{\l}odowska-Curie grant agreement 747645 (ClusterGal) and the Netherlands Organisation for Scientific Research (NWO) through VENI grant 639.041.751.




\bibliographystyle{mnras}
\bibliography{/Users/chessipearce/Documents/PhD/Bibtex/metallicity.bib} 



\appendix
\section{True vs. spectroscopic data}\label{ap:comp}

In order to compare to observations, it is common for simulation based studies to try and produce mock data which more closely resembles observational data. In this paper, we compare the results from the simulations directly to the observations as we have found that using mock spectroscopic data does not change the results of the paper. Here, we compare to \cite{Barnes2017b} who used mock spectroscopic data in their analysis of the C-EAGLE clusters. The spec data was produced using the method of \cite{LeBrun2014}, in which a rest frame X-ray spectrum in the band 0.5-10.0~keV was produced for every gas particle with $T > 10^{5}$~K, using the Astrophysical Plasma Emission Code (\textsc{APEC}, \citeauthor{Smith2001} \citeyear{Smith2001}) via the \textsc{PyAtomDB} module with atomic data from AtomDB 3.0.8 \citep{Foster2012} which includes emission lines. Individual spectra were created for all of the 11 elements tracked by the simulations: H, He, C, N, O, Ne, Mg, Si, S, Ca and Fe. Particles were separated into logarithmic radial bins, and the summed spectra within each bin was scaled by the relative abundance of heavy elements as the fiducial spectra assume solar abundance \citep{Anders1989}. A single temperature APEC model was fitted to the spectrum in the range 0.5-10.0~keV to give an estimate of the temperature, density and metallicity in each radial bin. During fitting, the spectrum in each energy bin was multiplied by the effective area of \textit{Chandra} to provide a better match to X-ray observations. This method does not fully reproduce X-ray observational analyses as the effects of projecting the data are not considered.

\begin{figure}
	\includegraphics[width=\columnwidth,keepaspectratio=True,trim=0cm 1cm 0cm 0cm]{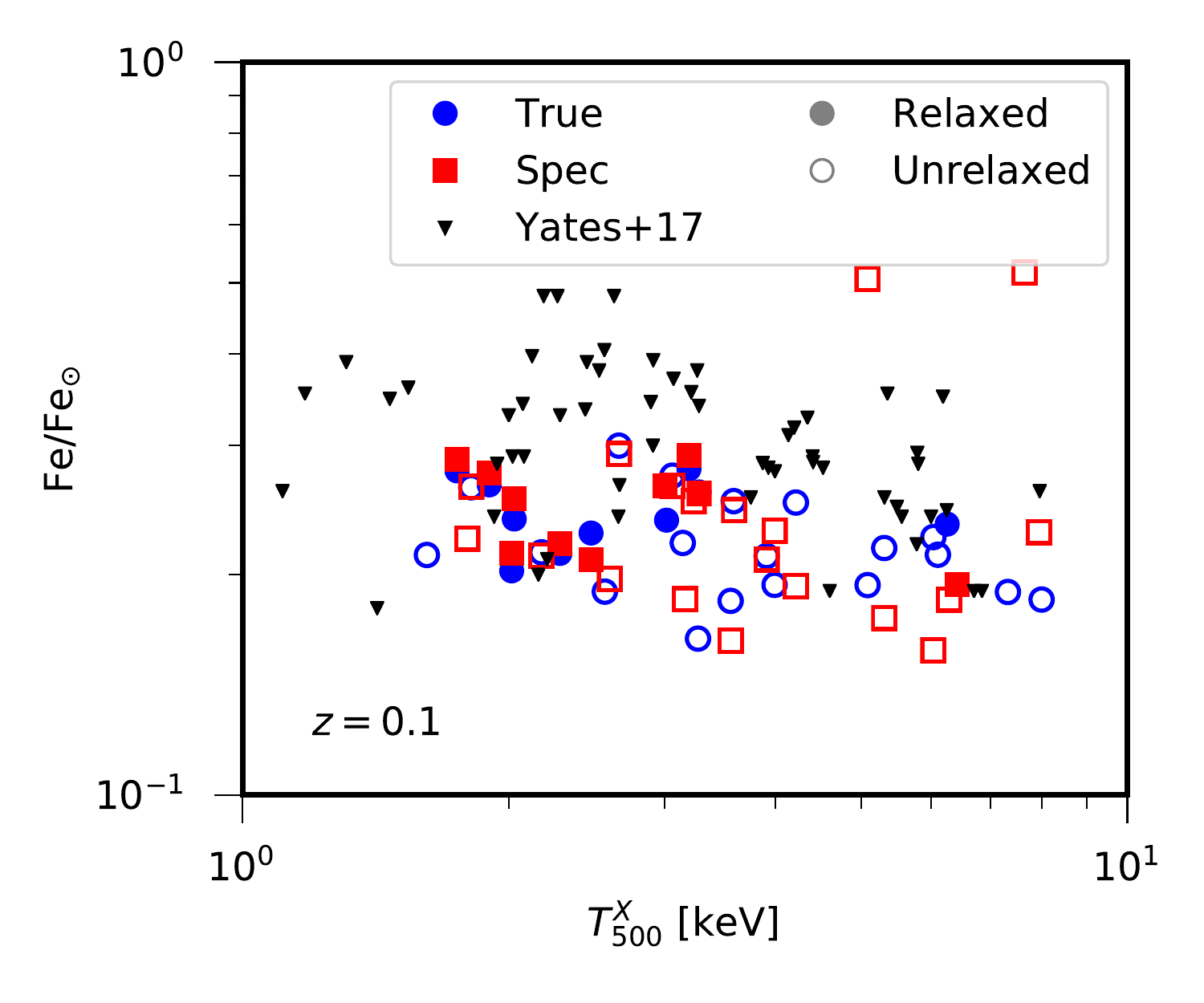}
	\caption{A comparison of the metallicity - temperature relation for the C-EAGLE clusters when using the true simulation data (blue circles), as in this paper, and the mock spectroscopic data (red squares) used in \citet{Barnes2017b}. Here we have partially reproduced fig.~9 of \citet{Barnes2017b}, using the solar abundances of \citet{Asplund2009} and presenting the data at $z=0.1$. The black triangles show the consolidated observational data from \citet{Yates2017} which falls within the plotted temperature range.}
	\label{fig:Fe_T}
\end{figure}

Fig.~\ref{fig:Fe_T} partly reproduces fig.~9 of \cite{Barnes2017b} and shows the metallicity - temperature relation for the 29 C-EAGLE clusters considered in this paper. The spec data used by \cite{Barnes2017b} is shown as red squares while the true simulation data used in this analysis is give by the blue circles. (Un)Relaxed clusters are depicted as (open)filled data points, and the combined observational sample of \cite{Yates2017} is given by the black triangles. Here we use the solar abundance of \cite{Asplund2009} and present the data at $z=0.1$, following \cite{Barnes2017b}.

Comparing the spec and true data, it can be seen that the temperature and Fe abundance of the individual clusters change but in general there is good agreement between the datasets. The spec data has a median metallicity of $Fe_{500,\mathrm{spec}} = 0.23$~Fe$_{\odot}$ with an r.m.s scatter of $\sigma = 0.08$. These values are different from those quoted by \cite{Barnes2017b} due to the removal of CE-27 which evolves unphysically due to a large AGN feedback event at very high redshift. Similarly, the true data has median metallicity $Fe_{500,\mathrm{true}} = 0.22$~Fe$_{\odot}$ with an r.m.s scatter of $\sigma = 0.03$, where the scatter is smaller than the spec data due to the decrease in metallicity of the clusters at 5~keV and 8~keV when moving to the true data. Both the true and the spec datasets are too low compared to the observations of \cite{Yates2017}, which is due to the C-EAGLE simulations not producing enough Fe (see Fig.~\ref{fig:Z_Fe_comparison}).

\begin{figure}
	\includegraphics[width=\columnwidth,keepaspectratio=True,trim=0cm 1cm 0cm 0cm]{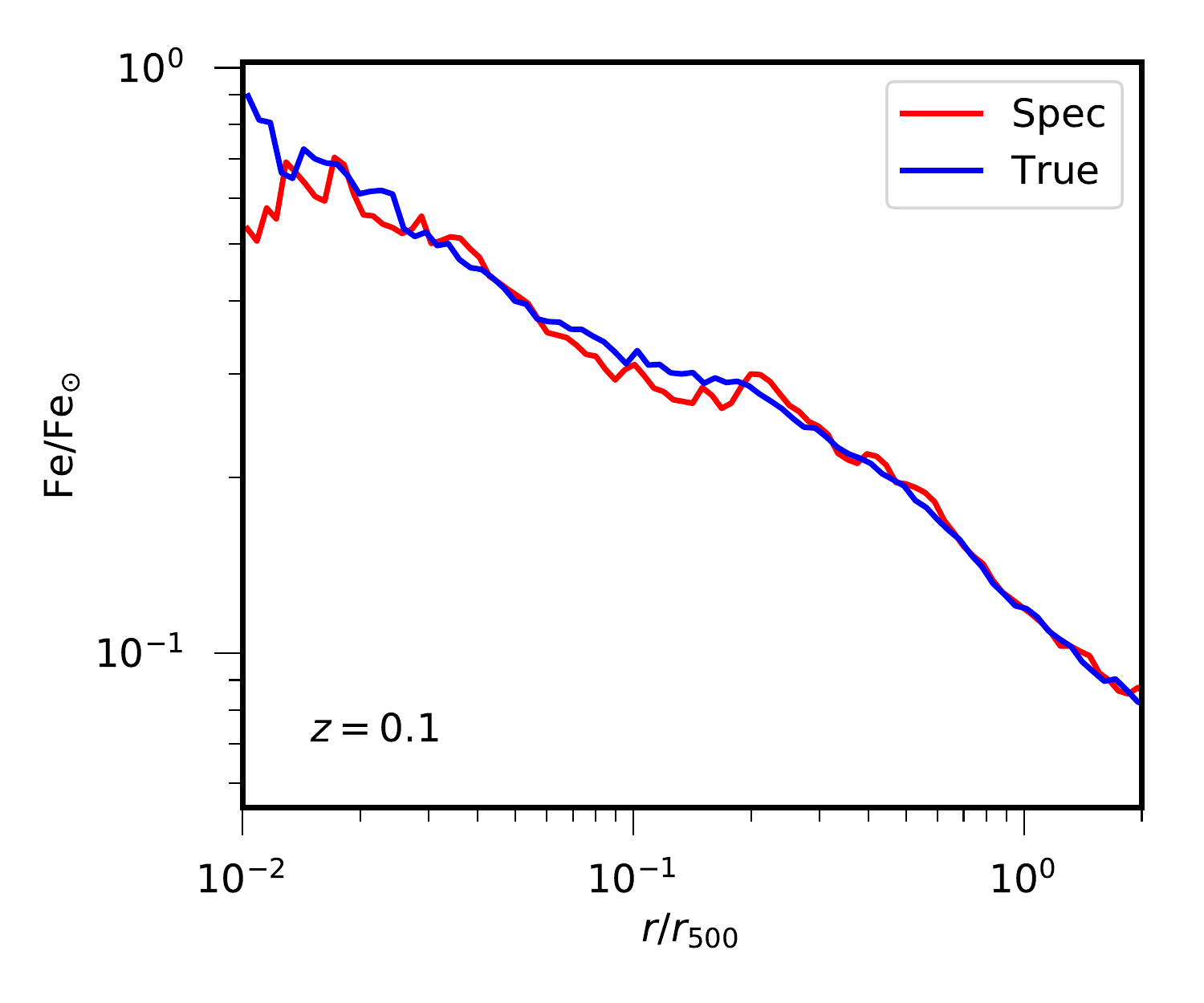}
	\caption{A comparison of the median Fe profiles obtained when using the data directly from the C-EAGLE simulations (blue line) and when using the method of \citet{LeBrun2014} to produce mock spectroscopic data (red line). We present the profiles at $z=0.1$ and use the solar abundances of \citet{Asplund2009} to compare directly to the work already done by \citet{Barnes2017b}.}
	\label{fig:profile_comp}
\end{figure}

Fig.~\ref{fig:profile_comp} compares the Fe abundance profiles from this analysis (true, blue) and \cite{Barnes2017b} (spec, red). Again we use the solar abundances of \cite{Asplund2009} and present the profiles at $z=0.1$. The spec profile is only presented out to 3~$r_{500}$ compared to 7~$r_{500}$ for the true profile, but in general there is good agreement between the profiles. 


\bsp	
\label{lastpage}
\end{document}